\documentclass[journal,draftclsnofoot,onecolumn,12pt]{IEEEtran}
\IEEEoverridecommandlockouts

\usepackage{acronym}
\usepackage{amsfonts}
\usepackage[dvips]{graphicx}
\usepackage{times}
\usepackage{cite}
\usepackage{amsmath}
\usepackage{array}
\usepackage{amssymb}
\usepackage{stfloats}
\usepackage{diagbox}
\usepackage{graphicx}
\usepackage{footnote}
\usepackage{amsthm}
\usepackage{booktabs}
\usepackage{array}
\usepackage[ruled,vlined]{algorithm2e}
\usepackage{subeqnarray}
\usepackage{cases}
\usepackage{threeparttable}
\usepackage{color}
\usepackage{epstopdf}
\usepackage{multirow}
\usepackage{tabularx}
\usepackage{enumerate}
\usepackage{multicol}
\usepackage{hyperref}
\usepackage{subfig}
\usepackage{caption}

\newtheorem{proposition}{Proposition}

\def\bb0{{\mathbb{0}}}

\def\bb{{\mathbf{b}}}

\def\bee{{\mathbf{e}}}

\def\bh{{\mathbf{h}}}

\def\bn{{\mathbf{n}}}

\def\bp{{\mathbf{p}}}

\def\bv{{\mathbf{v}}}

\def\bx{{\mathbf{x}}}
\def\by{{\mathbf{y}}}
\def\bz{{\mathbf{z}}}
\def\b0{{\mathbf{0}}}

\def\bA{{\mathbf{A}}}

\def\bD{{\mathbf{D}}}
\def\bE{{\mathbf{E}}}
\def\bF{{\mathbf{F}}}

\def\bH{{\mathbf{H}}}
\def\bI{{\mathbf{I}}}

\def\bN{{\mathbf{N}}}

\def\bR{{\mathbf{R}}}

\def\bV{{\mathbf{V}}}

\def\bX{{\mathbf{X}}}
\def\bY{{\mathbf{Y}}}

\def\bbC{{\mathbb{C}}}

\def\cF{\mathcal{F}}

\def\sf0{{\mathsf{0}}}

\def\rmA{\mathrm{A}}
\def\rmB{\mathrm{B}}

\def\rmD{\mathrm{D}}

\def\rmd{{\mathrm{d}}}

\def\rmp{{\mathrm{p}}}

\def\rm0{{\mathrm{0}}}

\acrodef{CSI}[CSI]{channel state information}
\acrodef{CSIT}[CSIT]{channel state information at the transmitter}
\acrodef{CSIR}[CSIR]{channel state information at the receiver}
\acrodef{MIMO}[MIMO]{multiple-input multiple-output}
\acrodef{SISO}[SISO]{single-input single-output}
\acrodef{MISO}[MISO]{multiple-input single-output}
\acrodef{SIMO}[SIMO]{single-input multiple-output}
\acrodef{ADCs}[ADCs]{analog-to-digital convertors}
\acrodef{SNR}[SNR]{signal-to-noise ratio}
\acrodef{AWGN}[AWGN]{additive white Gaussian noise}
\acrodef{MRT}[MRT]{maximal ratio transmission}
\acrodef{DFT}[DFT]{Discrete Fourier Transform}
\acrodef{ULA}[ULA]{uniform linear array}
\acrodef{UPA}[UPA]{uniform planar array}
\acrodef{LS}[LS]{least squares}
\acrodef{ALMMSE}[ALMMSE]{approximate linear minimum mean squared error}
\acrodef{QIHT}[QIHT]{quantized iterative hard thresholding}
\acrodef{QIST}[QIST]{quantized iterative soft thresholding}
\acrodef{SVD}[SVD]{singular value decomposition}

\usepackage{makecell}
\newcommand{\blc}[1]{\color{blue}#1}

\ifCLASSINFOpdf

\else
\fi
\begin{document}
\title{Cell-Free Massive MIMO Detection: A Distributed Expectation Propagation Approach}
\author{Hengtao He,~\IEEEmembership{Member,~IEEE,}
Xianghao Yu,~\IEEEmembership{Member,~IEEE,}
Jun Zhang,~\IEEEmembership{Fellow,~IEEE,}
S.H. Song,~\IEEEmembership{Senior,~Member,~IEEE,}
and Khaled B. Letaief,~\IEEEmembership{Fellow,~IEEE}
\thanks{This paper has been presented in part at the 2021 \emph{IEEE Global Commun. Conf. (GLOBECOM)}, Madrid, Spain. \cite{Distributed-EP}. This work was supported by the Research Grant Council under Grant No. 16212120.}
\thanks{H. He, X. Yu, J. Zhang, S. Song and Khaled B. Letaief are with the Department of Electronic and Computer Engineering,
the Hong Kong University of Science and Technology, Hong Kong, E-mail: \{eehthe, eexyu, eejzhang, eeshsong, eekhaled@ust.hk\}@ust.hk.}

}
\maketitle
\begin{abstract}
Cell-free massive MIMO is  one of the core technologies for future wireless  networks.  It is expected to bring enormous benefits, including ultra-high reliability, data throughput, energy efficiency, and uniform coverage.  As a radically distributed system, the performance of cell-free massive MIMO critically relies on efficient distributed processing algorithms. In this paper, we propose a distributed expectation propagation (EP) detector for cell-free massive MIMO, which consists of two modules: a nonlinear module at the central processing unit (CPU) and a linear module at each access point (AP). The turbo principle in iterative channel decoding is utilized to compute and pass the extrinsic information between the two modules. An analytical framework is provided to characterize the asymptotic performance of the proposed EP detector with a large number of antennas. Furthermore, a distributed iterative channel estimation and data detection (ICD) algorithm is developed to handle the practical setting with imperfect channel state information (CSI). Simulation results will show that the proposed method outperforms existing detectors for cell-free massive MIMO systems in terms of the  bit-error rate and demonstrate that the developed theoretical analysis accurately predicts system performance. Finally, it is shown that with imperfect CSI, the proposed ICD algorithm improves the system performance significantly and enables non-orthogonal pilots to reduce the pilot overhead.
\end{abstract}


%
\IEEEpeerreviewmaketitle

\section{Introduction}
The fifth-generation (5G) wireless networks have been commercialized since 2019 to support a wide range of services, including enhanced mobile broadband, ultra-reliable and low-latency communications, and massive machine-type communications\cite{5G}. However, the endeavor for faster and more reliable wireless communications will never stop. This trend is reinforced by the recent emergence of several innovative applications, including the Internet of Everything, Tactile Internet, and seamless virtual and augmented reality. Future sixth-generation (6G) wireless  networks are expected to provide ubiquitous coverage, enhanced spectral efficiency (SE), connected intelligence, etc. \cite{Khlaed_JSAC,6G_khaled}. Such diverse service requirements  create daunting challenges for system design. With the commercialization of  massive MIMO technologies \cite{massiveMIMO} in 5G, it is time to think about new MIMO-based network architectures to support the continuous exponential growth of mobile data traffic and a plethora of applications.

As a promising solution, cell-free massive MIMO was proposed \cite{cell_free}. It is a disruptive technology and has been recognized as a crucial and core enabler for beyond 5G and 6G networks\cite{cell_free_Zhang, beyond5G, 6G_matthaiou,JCIN_He, cell_implement,Ubiquitous_cell}. Cell-free massive MIMO can be interpreted as a combination of massive MIMO\cite{massiveMIMO}, distributed antenna systems (DAS)\cite{DAS}, and Network MIMO\cite{Zhang_Network}. It is expected to bring important benefits, including huge data throughput, ultra-low latency, ultra-high reliability, a huge increase in the mobile energy efficiency, and ubiquitous and uniform coverage.  In cell-free massive MIMO systems, a very large number of distributed access points (APs) are connected to a central processing unit (CPU) via a front-haul network in order to cooperate and jointly serve a large number of users using the same time or frequency resources over a wide coverage area.  In contrast to current cellular systems, there is no cell or cell boundary in cell-free MIMO networks.  As a result, this approach is revolutionary and will be able to relieve one of the major bottlenecks and inherent limitations of wireless networks, i.e., the strong inter-cell interference. Compared to conventional co-located massive MIMO, cell-free networks offer more uniform connectivity for all users thanks to the macro-diversity gain provided by the distributed antennas.

Investigations on cell-free massive MIMO started with some initial attempts on analyzing the SE \cite{cell_free}, where single-antenna APs, single-antenna users, and Rayleigh fading channels were considered. The analysis has been extended to multi-antenna APs with Rayleigh fading, Rician fading, and correlated channels \cite{performance_analysis1,performance_analysis2,performance_analysis3}, showing that cell-free massive MIMO networks can achieve great SE. The energy efficiency of cell-free massive MIMO systems was then investigated \cite{performance_analysis6,performance_analysis8}. It was shown that cell-free massive MIMO systems can improve the energy efficiency by approximately ten times compared to cellular massive MIMO.  Although cell-free massive MIMO has shown a huge potential, its deployment critically depends on effective and scalable algorithms. 
According to \cite{scalable},  cell-free massive MIMO is considered to be scalable if the signal processing tasks for channel estimation, precoder and combiner design, fronthaul overhead, and power optimization per AP can be kept within finite complexity when the number of served users goes to infinity. Unfortunately, the conventional cell-free massive MIMO is not scalable. To tackle the scalability issue, the concept of {\em user-centric} has been introduced into cell-free massive MIMO systems \cite{scalable,DCC1,DCC2,DCC3,DCC4,DCC5}. In particular, a user-centric dynamic cooperation clustering (DCC) scheme was introduced \cite{scalable}, where each user is only served by a subset of APs. This scheme was called {\em scalable} cell-free massive MIMO.

In this paper, we will focus on effective data detection algorithms in cell-free massive MIMO systems.
In this aspect, some early attempts were made on centralized algorithms where the detection is  implemented at the CPU with the received pilots and data signals reported from all APs \cite{cell_free, LSFD}. However, the computational and fronthaul overhead of such a centralized detection scheme is prohibitively high when the network size becomes large. To address this challenge, distributed detectors have been recently investigated. In \cite{ cell_free_distributed_receiver}, one centralized and three distributed receivers with different levels of cooperation among APs were compared in terms of SE. Radio stripes were then incorporated into cell-free massive MIMO in \cite{MMSE_optimal}.  In this case, the APs are sequentially connected and share the same fronthaul link in a radio stripe network, thus reducing the cabling substantially. Based on this structure, a novel uplink sequential processing algorithm was developed which can achieve an optimal performance in terms of both SE and mean-square error (MSE). Furthermore, it can achieve the same performance as the centralized minimum MSE (MMSE) processing, while requiring much lower fronthaul overhead by making full use of the computational resources at the APs. However, the distributed detectors investigated in \cite{cell_free_distributed_receiver} are \emph{linear} detectors. Therefore, they are highly suboptimal or may even be ill-conditioned in terms of the bit-error rate (BER) performance. To address this problem, the local per-bit soft detection is exploited at each AP  with the bit log-likelihoods  shared on the front-haul link by a partial marginalization detector \cite{soft_detection}. However, the proposed soft detection is still very complex as the approximate posterior density for each received data bit is required to be computed at each AP. Therefore, it is of great importance to develop a distributed  and \emph{non-linear} receiver to achieve a better BER performance with a considerably lower complexity.

To fill this gap, we propose a distributed non-linear detector for cell-free massive MIMO networks in this paper, which is derived based on the expectation propagation (EP) principle \cite{EP_principle}. The EP algorithm, proposed in \cite{EP_principle}, provides an iterative method to recover the transmitted data from the received signal and has recently attracted extensive research interests in massive MIMO detection \cite{EP_detector, Decentralized-AMP, EP_distributed, EP_imperfect_CSI}. {\blc Recently, it has been extended to cell-free massive MIMO systems \cite{EP_semi_blind,EP_Cell_free}.} The algorithm is derived from the factor graph with the messages updated and passed between different pairs of nodes. Specifically, with the linear MMSE estimator, the APs first detect the symbols with the local channel state information (CSI) and transfer the posterior mean and variance estimates to the CPU. Then, the extrinsic information for each AP is computed and integrated at the CPU by maximum-ratio combining (MRC). Subsequently, the CPU uses the posterior mean estimator to refine the detection and the extrinsic information is transferred to each AP from the CPU via the fronthaul.

The main contributions of this paper are summarized as follows:
\begin{itemize}
  \item Different from the existing linear detectors in cell-free massive MIMO\cite{cell_free_distributed_receiver}, we propose a distributed and non-linear detector.  The detection performance is improved at the cost of slightly increasing the computation overhead  at the computationally-powerful CPU. Simulation results will demonstrate that the proposed method outperforms existing distributed detectors and even the centralized MMSE detector in terms of BER performance.
   \item To be applicable in practical scenarios with imperfect CSI, we further develop a distributed  iterative  channel estimation and data detection (ICD) algorithm for cell-free massive MIMO systems, where the detector takes the channel estimation error and channel statistics into consideration while the channel estimation is refined by the detected data. Simulation results will demonstrate that the proposed ICD algorithm outperforms existing distributed detectors and enables non-orthogonal pilots to reduce the pilot overhead.
  \item We develop an analytical framework to analyze the performance of the distributed EP algorithm, which can precisely predict the performance of the proposed detector. Based on the theoretical analysis, key performance metrics of the system, such as the MSE  and BER, can be analytically determined without time-consuming Monte Carlo simulation.
\end{itemize}

\emph{Notations}---For any matrix $\mathbf{A}$, $\mathbf{A}^{H}$ and ${\mathrm{tr}}(\mathbf{A})$ will denote the conjugate transpose and  trace of $\mathbf{A}$, respectively. In addition, $\mathbf{I}$ is the identity matrix and $\mathbf{0}$ is the zero matrix.
We use $\rmD z$ to denote the real Gaussian integration measure. That is,
\begin{equation*}
  \rmD z=\phi(z)dz, \quad \mathrm{where} \quad  \phi(z)\triangleq\frac{1}{\sqrt{2\pi}}e^{-\frac{z^{2}}{2}}.
\end{equation*}
A complex Gaussian distribution with mean $\boldsymbol{\mu}$ and covariance $\boldsymbol{\Omega}$ can be described by the probability density function,
\begin{equation*}
  \mathcal{N}_{\mathbb{C}}(\mathbf{z};\boldsymbol{\mu},\boldsymbol{\Omega})=\frac{1}{\mathrm{det}(\pi \boldsymbol{\Omega})}
  e^{-(\mathbf{z}-\boldsymbol{\mu})^{H}\boldsymbol{\Omega}^{-1}(\mathbf{z}-\boldsymbol{\mu})}.
\end{equation*}

The remaining part of this paper is organized as follows. Section \ref{Problem} introduces the system model and formulates the cell-free massive MIMO detection problem. The distributed EP detector is proposed in Section \ref{SEC:EP} and the distributed ICD receiver is investigated in Section \ref{SEC:ICD}. Furthermore, an analytical framework is provided in Section \ref{SE:performance}. Numerical results are then presented in Section \ref{simulation} and Section \ref{con} concludes the paper.
\section{System Model}\label{Problem}
In this section, we first present the system model and formulate the cell-free massive MIMO detection problem. Four commonly-adopted receivers  are then briefly introduced.
\subsection{Cell-Free Massive MIMO}
As illustrated in Fig.\,\ref{cell_free}, we consider a cell-free massive MIMO network with $L$ distributed APs, each equipped with $N$ antennas to serve  single-antenna users. 
All APs are connected to a CPU that has abundant computing resources. Denote $\bh_{kl} \in \bbC^{N\times 1} \sim \mathcal{N}_{\bbC}(\mathbf{0}, \bR_{kl})$ as the channel between the $k$-th user and the $l$-th AP, where $\bR_{kl}\in \bbC^{N\times N}$ is the spatial correlation matrix, and $\beta_{k,l} = \mathrm{tr}(\bR_{kl})/N$ as the large-scale fading coefficient involving the geometric path loss and shadowing.  In the uplink data transmission phase, we consider $\mathcal{M}_{k} \subset \{1,\ldots,L\}$ as the subset of APs that serve the $k$-th user and define the DCC matrices $\bD_{kl} \in \bbC^{N\times N}$ based on $\mathcal{M}_{k}$ as
\begin{equation}\label{eqDCC}
  \bD_{kl} = \left\{
  \begin{aligned}
  &\bI_{N} \quad  \mathrm{if} \ l \in \mathcal{M}_{k}   \\
  & \mathbf{0}_{N\times N} \quad  \mathrm{if} \ l \notin \mathcal{M}_{k}.
  \end{aligned} \right.
\end{equation}
Specifically, $\bD_{kl}$ is an identity matrix if the  $l$-th AP serves the  $k$-th user, and is a zero matrix, otherwise.
Furthermore, we define $\mathcal{D}_{l}$ as the set of user indices that are served by the $l$-th AP
\begin{equation}\label{eqDl}
  \mathcal{D}_{l} = \bigg\{k: \mathrm{tr}(\bD_{kl}) \geq 1, k\in \{1,\ldots,K \} \bigg\},
\end{equation}
where the cardinality of $\mathcal{D}_{l}$ is denoted as $|\mathcal{D}_{l}|$. When each AP serves all users, we have
$|\mathcal{D}_{l}| = K$ accordingly. Assuming that perfect CSI is available at the local APs, the received signal $\by_{l} \in \bbC^{N\times 1}$ at the $l$-th AP is then given by
\begin{equation}\label{eqyd}
\by_{l} = \sum_{k\in\mathcal{D}_{l}}\sqrt{p_k}\bh_{kl}x_{k}+\bn_{l},
\end{equation}
where $x_{k}\in \bbC$ is the transmitted symbol drawn from an $M$-QAM constellation, and $p_{k}>0$ is the transmit power at the $k$-th user. The additive noise at the $l$-th AP is denoted as $\bn_{l} \in \bbC^{N\times 1} \sim \mathcal{N}_{\bbC}(\mathbf{0}_{N},\sigma^{2}\bI_{N})$. Let $\bx=[x_{1},\ldots,x_{K}]^{T}$ denote the transmitted vector from all users and  $\bH_{l}=[\bh_{1l},\ldots,\bh_{Kl}] \in \bbC^{N \times K}$  is the channel  from all users to  AP $l$. If the $k$-th user is not associated with the $l$-th AP, the channel vector is $\bh_{kl} = \mathbf{0}$. Furthermore, we denote   $\bH=[\bH_{1}^{T},\ldots,\bH_{L}^{T}]^{T}\in\bbC^{LN \times K}$ as the channel matrix between all users and APs. The uplink detection problem for cell-free massive MIMO is to detect the transmitted data $\bx$ based on the received signals $\by_{l}\,(l=1,2,\ldots,L)$, channel matrix $\bH$, and noise power $\sigma^{2}$.
\begin{figure*}
\begin{minipage}{3.5in}
  \centerline{\includegraphics[width=3.8in]{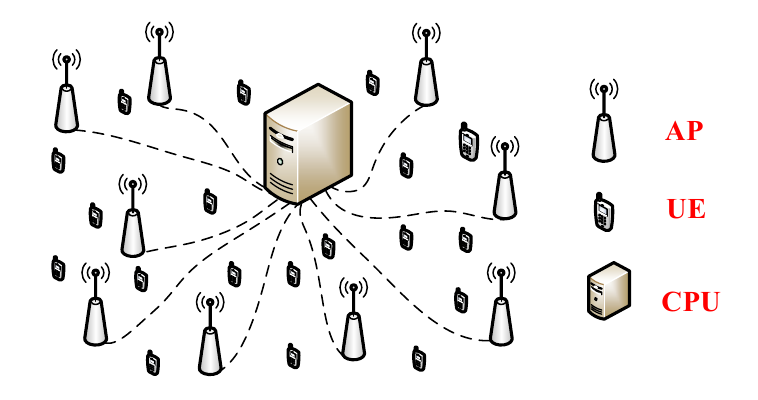}}
  \centerline{(a) Conventional cell-free massive MIMO system}\label{cell_free}
\end{minipage}
\hfill
\begin{minipage}{3.5in}
  \centerline{\includegraphics[width=3.8in]{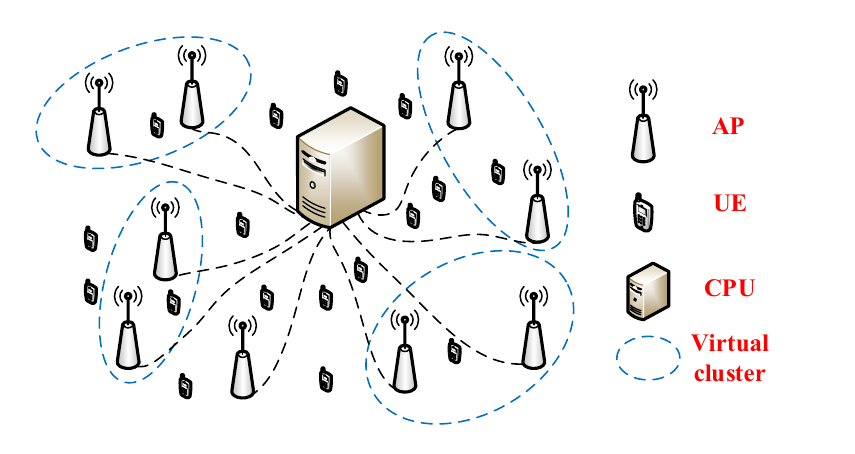}}
  \centerline{(b) Scalable cell-free massive MIMO system.} \label{cell_free_DCC}
\end{minipage}
\caption{.~~Conventional and scalable cell-free massive MIMO systems.}
\label{cell_free}
\end{figure*}
\subsection{Linear Receivers}
For the cell-free massive MIMO detection, there are four commonly-adopted linear receivers \cite{cell_free_distributed_receiver} with different levels of cooperation among APs.  This is  elaborated as follows.
\begin{itemize}
  \item \emph{Fully centralized receiver}: The pilot and data signals received at all APs are sent to the CPU for channel estimation and data detection. The CPU may perform the MMSE or MRC detection.
  \item \emph{Partially distributed receiver with large-scale fading decoding}: First, each AP estimates the channels and uses the linear MMSE detector to detect the received signals. Then, the detected signals are collected at the CPU for joint detection for all users by utilizing the large-scale fading decoding (LSFD) method. Compared to the fully centralized receiver, only the channel statistics are utilized at the CPU but the pilot signals are not required to be sent to the CPU.
  \item \emph{Partially distributed receiver with average decoding}: It is a special case of the partially distributed receiver with LSFD. The CPU performs joint detection for all users by simply taking the average of the local estimates. Thus, no channel statistics are required to be transmitted to the CPU via the fronthaul.
  \item \emph{Fully distributed receiver}: It is a fully distributed approach in which the data detection is performed  at the APs based on the local channel estimates. No information is required to be transferred to the CPU. Each AP may perform MMSE detection.
\end{itemize}

All of the above-mentioned four receivers achieve performance that is far from the optimal one due to the linear processing. In contrast, non-linear receivers have shown great advantages in terms of BER in cellular massive MIMO systems but at the cost of higher computational complexity \cite{EP_detector}. Thanks to the relatively high computing abilities at the CPU and the large number of APs in cell-free massive MIMO systems, we can offload some of the computational-intensive operations to the CPU and distribute the partial computation tasks to APs. Following this idea, we will propose a distributed non-linear detector for cell-free massive MIMO systems.
\section{Proposed Distributed EP Detector}\label{SEC:EP}
In this section, we first formulate the MIMO detection problem from the Bayesian inference. Then, we introduce the factor graph and propose the distributed EP detector. Finally, we will analyze the computational complexity and fronthaul overhead of the proposed detector.
\subsection{Distributed Bayesian Inference}
We first apply the Bayesian inference to recover the signals $\bx$ from the received signal $\by$ in the data detection stage with the linear model $\by = \bH \bx+\bn$, where $\by = [\by_1^T, \ldots, \by_L^T]^{T} \in \bbC^{LN\times 1}$, $\bH=[\bH_{1}^{T},\ldots,\bH_{L}^{T}]^{T}\in\bbC^{LN \times K}$ and $\bx=[x_{1},\ldots,x_{K}]^{T} \bbC^{K \times 1}$. Based on Bayes' theorem, the posterior probability is given by
\begin{equation}\label{eq4}
    \mathtt{P}(\bx|\by,\bH)=\frac{\mathtt{P}(\by|\bx,\bH)\mathtt{P}(\bx)}{\mathtt{P}(\by|\bH)}
 = \frac{\mathtt{P}(\by|\bx,\bH)\mathtt{P}(\bx)}{\int \mathtt{P}(\by|\bx,\bH)\mathtt{P}(\bx)d\bx}
\end{equation}
where $\mathtt{P}(\by|\bx,\bH)$ is the likelihood function with the known channel matrix and $\mathtt{P}(\bx)$ is the prior distribution of $\bx$.  Given the posterior probability $\mathtt{P}(\bx|\by,\bH)$, the Bayesian MMSE estimate is obtained by
\begin{equation}\label{MMSE_estimate}
\hat{\bx}=\int \bx \mathtt{P}(\bx|\by,\bH)d\bx.
\end{equation}
However, the Bayesian MMSE estimator is not tractable because the marginal posterior probability in
(\ref{MMSE_estimate}) involves a high-dimensional integral, which motivates us to develop an advanced method to approximate (\ref{eq4}) effectively. Furthermore, different from  conventional cellular MIMO systems\cite{EP_detector}, the posterior probability in (\ref{eq4}) has to be rewritten in a distributed way because of the inherent distributed characteristic of cell-free massive MIMO systems, given by
\begin{equation}\label{eqdistributed}
  \mathtt{P}(\bx|\by,\bH) \varpropto \mathtt{P}(\bx)\prod_{l=1}^{L} \mathrm{exp}(-\|\by_{l}-\bH_{l}\bx\|^{2}/\sigma^{2}).
\end{equation}
As a result, we can develop an efficient distributed MIMO detector to approximate the posterior probability (\ref{eqdistributed}) by resorting to message passing on factor graph and take the characteristic of cell-free massive MIMO into consideration. The factor graph and proposed detector will be elaborated  in next sections.
\subsection{Factor Graph for Cell-free Massive MIMO Detection}\label{Sec:message_passing}
\begin{figure}
  \centering
  \includegraphics[width = 6 in]{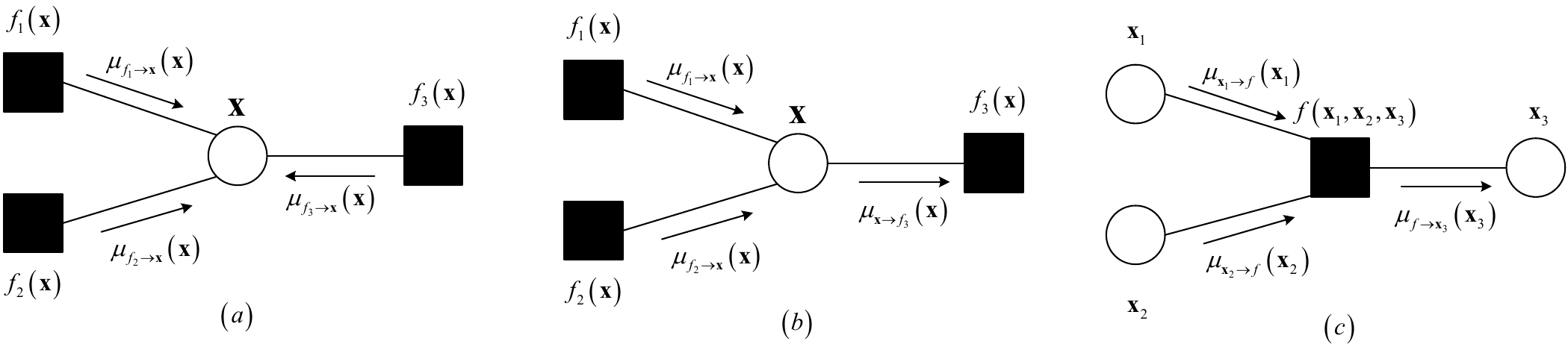}
  \caption{.~~Factor graphs for message passing algorithms.}\label{Fig:factor_graph}
\end{figure}

Factor graph is a powerful tool to develop efficient algorithms for solving inference problems by performing different message-passing rules. We first introduce several concepts for the factor graph \cite{factor_graph}, as illustrated in Fig.\,\ref{Fig:factor_graph}, where the hollow circles represent the variable nodes and the solid squares represent the factor nodes. For a factor graph with factor nodes $\{f_0,f_1,\ldots,f_L\}$, all connected to a set of variable nodes $\{\bx_i\}$, the messages are computed and updated iteratively by performing the following rules.

1) \emph{Approximate beliefs}: The approximate belief $b_\mathrm{app}(\bx)$ on variable node $\bx$ is $\mathcal{N}_{\mathbb{C}}(\bx,\bx_{\mathrm{A}}^{\mathrm{ext}},v_{\mathrm{A}}^{\mathrm{ext}})$, where $\bx_{\mathrm{A}}^{\mathrm{ext}}=\mathtt{E}[\bx|b_{\mathrm{sp}}]$ and $v_{\mathrm{A}}^{\mathrm{ext}} = \mathrm{mean}(\mathrm{diag}(\mathrm{Cov}[\bx|b_{\mathrm{sp}}]))$ are the mean and average variance of the corresponding beliefs $b_{\mathrm{sp}}(\bx) \propto \prod_l\mu_{f_{l}\rightarrow \bx}(\bx)$.

2) \emph{Variable-to-factor messages}: The message from a  variable node $\bx$ to a connected factor node $f_{l}$ is given by
  \begin{equation}\label{eqV2f}
    \mu_{\bx \rightarrow f_l}(\bx)= \prod_{j \neq l}^{L} \mu_{f_j \rightarrow \bx}(\bx),
  \end{equation}
which is the product of the message from other factors.

{\blc
3) \emph{Factor-to-variable messages}: The message from a factor node $f_l$ to a connected variable node $\bx_{i}$ is
$\mu_{f_l\rightarrow \bx_{i}}(\bx) \propto \int f(\bx_{i},\{\bx_{j}\}_{j\neq i})\prod_{j\neq i} \mu_{\bx_{j} \rightarrow f_l} \rmd\bx_{j}$.}

If we treat the $L$ APs and the CPU as the factor nodes, we can construct the factor graph for cell-free massive MIMO detection. By further applying the above message-passing rules to the constructed factor graph, we develop the distributed EP detector for cell-free massive MIMO systems. The detailed derivation is given in Appendix \ref{derivation}.
\begin{figure}
\begin{minipage}{3in}
  \centerline{\includegraphics[width=3.2in]{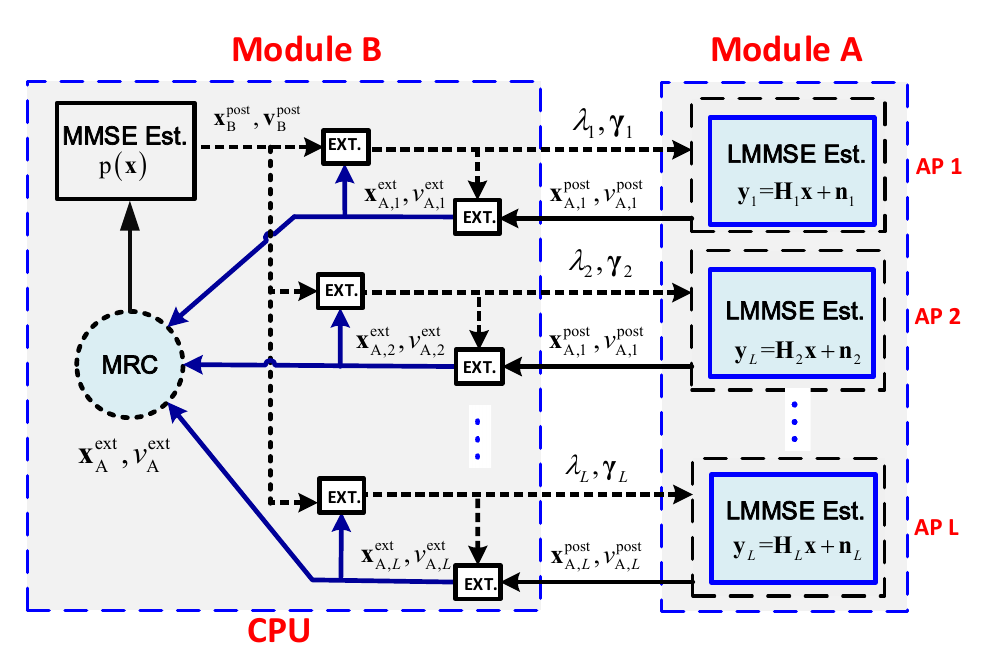}}
\end{minipage}
\hfill
\begin{minipage}{3in}
  \centerline{\includegraphics[width=3.6in]{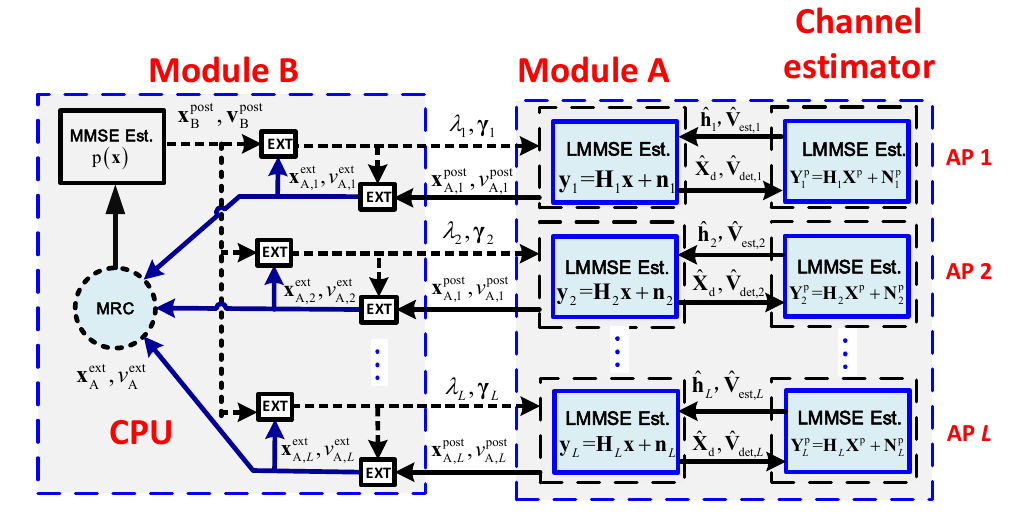}}
\end{minipage}
\caption{{\blc .~~ Block diagrams of the proposed distributed EP detector, as well as the distributed EP-based {\blc iterative} channel estimation and data detection. The ``EXT.'' blocks represent the extrinsic information computation. The channel estimator and signal detector exchange information iteratively until convergence.}}\label{DeEP}
\end{figure}
\subsection{Distributed EP Detector}
The distributed EP-based detector is illustrated in \textbf{Algorithm 1}. Compared with the centralized EP detector, it deploys partial calculation modules  at the AP based on the local information and then sends the estimated posterior mean and variance to the CPU for combining. In particular, the input of the algorithm is the received signal $\by_{l}$, channel matrix $\bH_{l}$, and noise level $\sigma^{2}$, while the output is the recovered signal $\bx_{\rmB}^{\mathrm{post},T}$ in the $T$-th iteration. The initial parameters are $\boldsymbol{\gamma}_{l}^{(0)} = \mathbf{0}$, and $\lambda_{l}^{(0)} = \frac{1}{E_{x}}$, where 
\begin{equation}\label{eqex}
E_{x}=\mathtt{E}\{\|\bx\|^{2}\}/K.
\end{equation}
Note that $\boldsymbol{\gamma}_{l}^{(0)}$ and $\lambda_{l}^{(0)}$ are the initialized extrinsic information, and $E_{x}$ is the power of the transmitted symbol $x_{k}$. The block diagram of the proposed distributed {\blc EP} detector is illustrated in {\blc Fig.\,\ref{DeEP}}, which is composed of two modules, A and B. Each module uses the turbo principle in iterative decoding, where each module passes the extrinsic messages to the other module and this process repeats until convergence.
It is derived from the factor graph with the messages updated and passed between different pairs of nodes that are assumed to follow Gaussian distributions. As the Gaussian distribution can be fully characterized by its mean and variance, only the mean and variance  need to be calculated and passed between different modules.
To better understand the distributed EP detection algorithm, we elaborate the details of each equation in \textbf{Algorithm 1}. Specifically, module A is the linear MMSE (LMMSE) estimator performed at the APs according to the following linear model
\begin{equation}\label{eqLMMSE}
  \by_{l} = \bH_{l}\bx+\bn_{l}.
\end{equation}
In the $t$-iteration, the explicit expressions for the posterior covariance matrix $\boldsymbol{\Sigma}_{l}^{t}$ and mean vector $\boldsymbol{\mu}_{l}^{t}$  are given by (\ref{eqvarA}) and (\ref{eqmeanA}), respectively. Note that each AP only uses the local channel $\bH_{l}$ to detect the transmitted signal $\bx$. For the ease of notation, we omit the iteration index $t$ for all estimates for the mean and variance in {\blc \textbf{Algorithm 1}}. Then, the variance estimate $v_{\rmA,l}^{\mathrm{post}} = \mathrm{tr}(\boldsymbol{\Sigma}_{l}^{t})/|\mathcal{D}_{l}|$ and mean estimate $\bx_{\rmA,l}^{\mathrm{post}} = \boldsymbol{\mu}_{l}^{t}$ are transferred to the CPU to compute the extrinsic information $v_{\rmA,l}^{\mathrm{ext}}$ (\ref{eqextvar}) and  $\bx_{\rmA,l}^{\mathrm{ext}}$ (\ref{eqextmean}), respectively. The extrinsic information $\bx_{\rmA,l}^{\mathrm{\mathrm{ext}}}$ can be regarded as the AWGN observation given by
\begin{equation}\label{eqAWGNl}
  \bx_{\rmA,l}^{\mathrm{ext}} = \bx+\bn^{\mathrm{ext}}_{\mathrm{A},l},
\end{equation}
where $\bn^{\mathrm{ext}}_{\mathrm{A},l}\sim \mathcal{N}_{\bbC} (0,v_{\rmA,l}^{\mathrm{ext}}\bI)$ \cite{EP_distributed}. As a result, the linear model in (\ref{eqyd}) is decoupled into $K$ parallel and independent AWGN channels with equivalent noise $v_{\rmA,l}^{\mathrm{ext}}$ for each AP. Subsequently, the CPU collects all extrinsic means $\{\bx_{\rmA,l}^{\mathrm{ext}}\}_{l=1}^{L}$ and variances  $\{v_{\rmA,l}^{\mathrm{ext}}\}_{l=1}^{L}$ and performs MRC\footnote{{\blc In our paper, we focus on linear combining of the local estimate $\mathbf{x}_{\mathrm{A},l}^{\mathrm{ext}}$ and  $v_{\mathrm{A},l}^{\mathrm{ext}}$ from APs as
$\mathbf{x}_{\mathrm{A}}^{\mathrm{ext}} = \sum_{l=1}^{L}a_{l}\mathbf{x}_{\mathrm{A},l}^{\mathrm{ext}}, \quad l = 1,\ldots,L $. The optimal combining is MRC, which
can maximize the  post-combination SNR of the final AWGN observation $\mathrm{x}_{\mathrm{A}}^{\mathrm{ext}}$  with the  linear combination of $\mathbf{x}_{\mathrm{A},l}^{\mathrm{ext}}$ where $\mathbf{x}_{\mathrm{A}}^{\mathrm{ext}} = \mathbf{x}+\mathbf{n}^{\mathrm{eq}}$ and $\mathbf{n}_{\mathrm{A}}^{\mathrm{ext}} \sim \mathcal{N}_{\mathbb{C}} (0,v_{\mathrm{A}}^{\mathrm{ext}}\mathbf{I})$. The final expressions are shown in (\ref{eqvarmrc}) and (\ref{eqmeanmrc}) and the  detailed derivation is given in Appendix \ref{derivation33}.}}. The MRC expressions (\ref{eqvarmrc}) and (\ref{eqmeanmrc}) are obtained by maximizing the post-combination signal-to-noise ratio (SNR) of the  AWGN observation $\bx_{\rmA}^{\mathrm{ext}}$  at the CPU, given by
{\blc
\begin{equation}\label{eqAWGN}
  \bx_{\rmA}^{\mathrm{ext}} = \bx+\bn^{\mathrm{ext}}_{\mathrm{A}},
\end{equation}
where $\bn^{\mathrm{ext}}_{\mathrm{A}} \sim \mathcal{N}_{\bbC} (0,v_{\rmA}^{\mathrm{ext}}\bI)$.} The CPU uses the posterior mean estimator to detect the signal $\bx$ from the equivalent AWGN model (\ref{eqAWGN}). Then,
the posterior mean and variance are computed by the posterior MMSE estimator for the equivalent AWGN model in (\ref{eqmeanB}) and (\ref{eqvarB}).
As the transmitted symbol is assumed to be drawn from the $M$-QAM set $\mathcal{S}=\{s_{1}, s_{2}, \ldots, s_{M}\}$, the corresponding expressions for each element in (\ref{eqmeanB}) and (\ref{eqvarB}) are given by
\begin{equation}\label{eqmean}
  [x_{\rmB}^{\mathrm{post}}]_{k} =\mathtt{E} \{x_k|[x_{\mathrm{A}}^{\mathrm{ext}}]_{k}, v_{\mathrm{A}}^{\mathrm{ext}}\} =\frac{\sum_{s_{i} \in \mathcal{S}}s_{i}\mathcal{N}_{\bbC}(s_{i};[x_{\rmA}^{\mathrm{ext}}]_{k}, v_{\rmA}^{\mathrm{ext}})p(s_{i})}{\sum_{s_{i} \in \mathcal{S}}\mathcal{N}_{\bbC}(s_{i};[x_{\rmA}^{\mathrm{ext}}]_{k}, v_{\rmA}^{\mathrm{ext}})p(s_{i})},
\end{equation}
\begin{equation}\label{eqvar}
  [v_{\rmB}^{\mathrm{post}}]_{k} = \mathtt{var} \{x_k|[x_{\mathrm{A}}^{\mathrm{ext}}]_{k}, v_{\mathrm{A}}^{\mathrm{ext}}\} = \frac{\sum_{s_{i} \in \mathcal{S}}|s_{i}|^{2}\mathcal{N}_{\bbC}(s_{i};[x_{\rmA}^{\mathrm{ext}}]_{k}, v_{\rmA}^{\mathrm{ext}})p(s_{i})}{\sum_{s_{i} \in \mathcal{S}}\mathcal{N}_{\bbC}(s_{i};[x_{\rmA}^{\mathrm{ext}}]_{k}, v_{\rmA}^{\mathrm{ext}})p(s_{i})}
  -|[x_{\rmB}^{\mathrm{post}}]_{k}|^{2},
\end{equation}
where $[x_{\rmB}^{\mathrm{post}}]_{k}$, $[v_{\rmB}^{\mathrm{post}}]_{k}$, and $[x_{\rmA}^{\mathrm{ext}}]_{k}$ are the $k$-th element in $\bx_{\rmB}^{\mathrm{post}}$, $\bv_{\rmB}^{\mathrm{post}}$, and $\bx_{\rmA}^{\mathrm{ext}}$, respectively. The posterior mean  $\bx_{\rmB}^{\mathrm{post}}$ and variance $\bv_{\rmB}^{\mathrm{post}}$ are then utilized to compute the extrinsic information $\lambda_{l}^{(t)}$ and $\boldsymbol{\gamma}_{l}^{(t)}$ for each AP in (\ref{eqvarest}) and (\ref{eqmmeanest}), where the function $\mathrm{mean}(\cdot)$ is used to compute the mean of a vector. Finally, the extrinsic information $\lambda_{l}^{(t)}$ and $\boldsymbol{\gamma}_{l}^{(t)}$ are transferred to each AP in the next iteration. The whole procedure is executed iteratively until it is terminated by a stopping criterion or a maximum number of iterations.
\begin{algorithm}\label{algGE}
\caption{Distributed EP for cell-free massive MIMO detection} 
{\bf Input:} 
Received signal $\by_{l}$, channel matrix $\bH_{l}$, and noise level $\sigma^{2}$. \\
{\bf Output:} 
Recovered signal $\bx_{\rmB}^{\mathrm{post}}$.\\
{\bf Initialize:}   
$\boldsymbol{\gamma}_{l}^{(0)} \leftarrow \mathbf{0}$, $\lambda_{l}^{(0)} \leftarrow \frac{1}{E_{x}}$ \\
\For{$t = 1,\cdots, T$ }{
\hspace*{0.02in} {\bf Module A in APs:}

(1) Compute the posterior mean and variance of $\bx_{\rmA,l}$:
\begin{equation}\label{eqvarA}
v_{\rmA,l}^{ \mathrm{post}} \leftarrow \boldsymbol{\Sigma}_{l}^{t} = \bigg(\sigma^{-2} \bH^{H}_{l}\bH_{l}+\lambda_{l}^{(t-1)}\bI\bigg)^{-1}
\end{equation}
\begin{equation}\label{eqmeanA}
\bx_{\rmA,l}^{\mathrm{post}} \leftarrow \boldsymbol{\mu}_{l}^{t} = \boldsymbol{\Sigma}_{l}^{t} \bigg( \sigma^{-2}\bH_{l}^{H}\by_{l}+
\boldsymbol{\gamma}_{l}^{(t-1)}\bigg).
\end{equation}
\hspace*{0.02in} {\bf Module B in CPU:}

(2) Compute the extrinsic mean and variance of $\bx_{\rmA,l}$:
    \begin{equation}\label{eqextvar}
     v_{\rmA,l}^{\mathrm{ext}} \leftarrow \bigg( \frac{1}{v_{\rmA,l}^{ \mathrm{post}}}-\lambda_{l}^{(t-1)}
      \bigg)^{-1}
    \end{equation}
\begin{equation}\label{eqextmean}
     \bx_{\rmA,l}^{\mathrm{ext}} \leftarrow v_{\rmA,l}^{\mathrm{ext}} \bigg(\frac{\boldsymbol{\mu}_{l}^{t}}{v_{\rmA,l}^{ \mathrm{post}}}-\boldsymbol{\gamma}_{l}^{(t-1)}
      \bigg)^{-1}.
\end{equation}
(3) MRC combining of $\bx_{\rmA,l}$:
\begin{equation}\label{eqvarmrc}
    \frac{1}{v_{\rmA}^{\mathrm{ext}}} = \sum_{l=1}^{L} \frac{1} {v_{\rmA,l}^{\mathrm{ext}}}
\end{equation}
\begin{equation}\label{eqmeanmrc}
    \bx_{\rmA}^{\mathrm{ext}} = v_{\rmA}^{\mathrm{ext}}\sum_{l=1}^{L}\frac{\bx_{\rmA,l}^{\mathrm{ext}}}{v_{\rmA,l}^{\mathrm{ext}}}.
\end{equation}
(4) Compute the posterior mean and variance of $\bx_{\rmB}$:
    \begin{equation}\label{eqmeanB}
    \bx_{\rmB}^{\mathrm{post}} = \mathtt{E} \{\bx|\bx_{\rmA}^{\mathrm{ext}}, v_{\rmA}^{\mathrm{ext}}\}
  \end{equation}
   \begin{equation}\label{eqvarB}
    \bv_{\rmB}^{\mathrm{post}} = \mathtt{var} \{\bx|\bx_{\rmA}^{\mathrm{ext}}, v_{\rmA}^{\mathrm{ext}}\}.
   \end{equation}
(5) Compute the extrinsic mean and variance $\bx_{\rmB,l}$:
\begin{equation}\label{eqvarest}
    \frac{1}{v_{\rmB,l}^{\mathrm{ext}}} \leftarrow \lambda_{l}^{(t)}
    = \frac{1}{\mathrm{mean}(\bv_{\rmB}^{\mathrm{post}})}-\frac{1}{v_{\rmA,l}^{\mathrm{ext}}}
\end{equation}
\begin{equation}\label{eqmmeanest}
  \frac{\bx_{\rmB,l}^{\mathrm{ext}}}{v_{\rmB,l}^{\mathrm{ext}}} \leftarrow \boldsymbol{\gamma}_{l}^{(t)}=    \frac{\bx_{\rmB}^{\mathrm{post}}}{\mathrm{mean}(\bv_{\rmB}^{\mathrm{post}})} -
  \frac{\bx_{\rmA,l}^{\mathrm{post}}}{v_{\rmA,l}^{\mathrm{ext}}}.
\end{equation}
}
\end{algorithm}

\subsection{Computational Complexity and Fronthaul Overhead}
{\blc In the following, we provide the complexity analysis and fronthaul overhead for different detectors in Tables \ref{tbl:complexity} and \ref{tbl:fronthaul}, respectively.} For the proposed distributed EP detector, the computational complexity at each AP is dominated by the LMMSE estimator for estimating the signal $\bx_{\mathrm{A},l}^{\mathrm{post}}$, which is $\mathcal{O}(N|\mathcal{D}_{l}|^{2})$ because of the matrix inversion  required  in (\ref{eqvarA}), while the computational complexity at the CPU is $O(L|\mathcal{D}_{l}|)$ in each iteration. Furthermore, if the number of antennas $N$ is less than $|\mathcal{D}_{l}|$, we can use the matrix inversion lemma to carry out the matrix inversion in (\ref{eqvarA}) as follows
\begin{equation}\label{eqmatrixinv}
   (\sigma^{-2}\bH_{l}^{H}\bH_{l}+\bD)^{-1} = \bD^{-1}  \nonumber -\sigma^{-2}\bD^{-1}\bH_{l}^{H}(\bI+\sigma^{-2}\bH_{l}\bD^{-1}\bH_{l}^{H})^{-1}\bH_{l}\bD^{-1},
\end{equation}
where $\bD = \lambda_{l}^{(t-1)}\bI$ and  the computational complexity is reduced to $\mathcal{O}(|\mathcal{D}_{l}| N^{2})$. Therefore, the overall computational complexity at each AP is $\mathcal{O}(T|\mathcal{D}_{l}|N^{2})$ while
the overall computational complexity at the CPU is $\mathcal{O}(TL|\mathcal{D}_{l}|)$ for $T$ iterations. As observed in Table \ref{tbl:complexity}, the distritbuted EP detector mainly increases the computational complexity at the CPU when compared with other distributed linear receivers.
{\blc
 \begin{table}[t]
	\centering
	\renewcommand{\arraystretch}{1.1}
	\begin{minipage}[c]{1\columnwidth}
		\centering
		\caption{.~~Complexity of different detectors.}
		\label{tbl:complexity}
		\begin{tabular}{@{}lcccc@{}}
			\toprule
			Detectors & Distributed EP & Centralized MMSE & Partially distributed MMSE & Distributed MMSE   \\
			\midrule
			  AP &
         \makecell[c]{  {\blc $8NT|\mathcal{D}_{l}|(|\mathcal{D}_{l}|+1)+$} \\ {\blc  $6T|\mathcal{D}_{l}|(|\mathcal{D}_{l}|+2)$  }     }&

             {\blc $0$} &

             \makecell[c]{   {\blc $8N|\mathcal{D}_{l}|(|\mathcal{D}_{l}|+1)+$} \\ {\blc $6|\mathcal{D}_{l}|(|\mathcal{D}_{l}|+2)$}  }&

             \makecell[c]{   {\blc $8N|\mathcal{D}_{l}|(|\mathcal{D}_{l}|+1)+$} \\ {\blc $6|\mathcal{D}_{l}|(|\mathcal{D}_{l}|+2)$} } \\	

			\midrule
              CPU &  \makecell[c]{  {\blc $LT(1+2|\mathcal{D}_{l}|)+$} \\   {\blc $|\mathcal{D}_{l}|(T-1)(7M+2)$ }    } &

              \makecell[c]{   {\blc $|\mathcal{D}_{l}|[8NL(|\mathcal{D}_{l}|+1) +$}   \\    {\blc $2(4|\mathcal{D}_{l}|+3)]$ }     } &

              {\blc $L(1+2|\mathcal{D}_{l}|)$}  &

              {\blc $0$} \\		
              				
			\bottomrule
		\end{tabular}

	\end{minipage}
\end{table}
}

We further compare the number of complex scalars that need to be transmitted from the APs to the CPU via the fronthauls of different detectors in Table \ref{tbl:fronthaul}. We assume that $\tau_c$ and $\tau_p$ are the coherence time and pilot length, respectively. The results for the 4 baseline detectors are from \cite{cell_free_distributed_receiver}. With the proposed detector, $\sum\limits_{l=1}^{L}(\tau_{c}-\tau_{p})2T(|\mathcal{D}_{l}|+1)$ scalars need to be passed from the APs to the CPU and no statistical parameters are required to be passed, where $T$ denotes the total number of iterations. {\blc The fronthaul overhead of the proposed distributed EP detector is similar to those of the  centralized MMSE and EP detectors. The detailed comparison shall be determined by the value of  system parameters.}
{\blc
\begin{table}[h]
	\centering
	\renewcommand{\arraystretch}{1.1}
	\begin{minipage}[c]{1\columnwidth}
		\centering
		\caption{.~~Fronthaul overhead}
		\label{tbl:fronthaul}
		\begin{tabular}{@{}lcc@{}}
		  \toprule
		   Detectors & Coherence block &  Statistical parameters  \\
		  \midrule
		  Distributed EP  &$\sum\limits_{l=1}^{L}(\tau_{c}-\tau_{p})2T(|\mathcal{D}_{l}|+1)$&$0$  \\
		  \midrule
		  Centralized MMSE  &$\tau_{c}NL$ &$KLN^{2}/2$ \\
		  \midrule
		  Partially distributed MMSE with LSFD  &$(\tau_{c}-\tau_{p})KL$&$KL+(L^{2}K^{2}+KL)/2$  \\
		  \midrule
		  Partially distributed MMSE with average decoding  &$(\tau_{c}-\tau_{p})KL$&$0$  \\
		  \midrule
		  Distributed MMSE  &$0$&$0$  \\
		  \midrule
		 {\blc Centralized EP}  &  {\blc$\tau_{c}NL$}        &      {\blc $KLN^{2}/2$ }   \\				
		 \bottomrule
		\end{tabular}
	\end{minipage}
\end{table}
}
\section{Distributed  iterative Channel Estimation and Data Detection}\label{SEC:ICD}
In Section \ref{SEC:EP}, we developed a distributed EP detector assuming perfect CSI. However, the channel matrix is typically estimated at the receiver with uplink training, and thus channel estimation errors should be considered in the detection stage. 
In this section, we will propose a distributed ICD receiver for cell-free massive MIMO to handle imperfect CSI.
\subsection{ICD Algorithm}
ICD has been explored for MIMO \cite{ICD_MIMO,OAMP-Net2}, OFDM \cite{ICD_OFDM}, and massive MIMO \cite{ICD_massive} systems with  a low-resolution  analog-to-digital converter (ADC) \cite{ICD_ADC}. Recently, it has  been investigated in cell-free massive MIMO to improve the  BER performance and reduce the pilot overhead. Different from \cite{ICD_cell_free} that solved the complex biconvex optimization problem with high complexity, we develop a low-complexity turbo-like ICD architecture. The iterative procedure is summarized in \textbf{Algorithm 2}.

As illustrated in Fig.\,\ref{DeEP}, the proposed ICD scheme employs a similar idea as iterative decoding. At each AP, the channel estimator and Module A in the proposed distributed EP detector are performed. They exchange extrinsic information iteratively. In the ICD processing stage, the channel estimation is first performed based on the received pilot signal and transmitted pilot. Then, the data detector performs signal detection by taking both the channel estimation error and channel statistics into consideration. It will then feed back the detected data and detection error to the channel estimator. Finally, data-aided channel estimation is employed with the help of the detected data. The whole ICD procedure is executed iteratively until it is terminated by a stopping criterion or a maximum number of ICD iterations.

Similar to \textbf{Algorithm 1}, the input of the ICD scheme is the pilot signal matrix, $\mathbf{X}^{\rmp}$, the received signal matrix corresponding to the pilot matrix, $\bY^{\rmp}$, and that corresponding to the data matrix, $\bY^{\rmd}$, in each time slot for each AP. In the $r$-th ICD iteration\footnote{ Here we use $r$ to denote the number of data feedback from the data detector to the channel estimator. For example, $r=1$ means no data feedback while $r=4$ means 3 data feedback should be performed.}, $\hat{\bH}^{(r)}$ is the estimated channel matrix, $\hat{\bX}^{(\rmd, r)}$ is the estimated data matrix, and $\hat{\bV}_{\mathrm{est},l}$ and $\hat{\bV}_{\mathrm{det}, l}$ are used to compute the covariance matrix for the equivalent noise in the signal detector and channel estimator, respectively. The final output of the signal detector is  the detected data matrix $\hat{\bX}^{(\mathrm{d},R)}$, where $R$ is the total number of ICD iterations. Compared with the conventional receiver  where the channel estimator and signal detector are designed separately, the ICD scheme can improve the BER performance by considering the characteristics of the channel estimation error and channel statistics. After performing signal detection, the detected payload data will be utilized for channel estimation. Next, we will elaborate the whole procedure in detail.
\begin{algorithm}\label{algGE}
\caption{Distributed EP-based ICD for cell-free massive MIMO} 
{\bf Input:} 
Received signal $\bY^{\rmp}$ and $\bY^{\rmd}$,  pilot signal matrix, $\mathbf{X}^{\rmp}$, spatial correlation matrix $\bR_{kl}$ , and noise level $\sigma^{2}$. \\
{\bf Output:} 
Recovered signal $\hat{\bX}_{(\mathrm{d}, R)} = \bx_{\rmB}^{\mathrm{post}}$.\\
{\bf Initialize:}   
$\boldsymbol{\gamma}_{l}^{(0)} \leftarrow \mathbf{0}$, $\lambda_{l}^{(0)} \leftarrow \frac{1}{E_{x}}$ \\

\For{$r = 1,\cdots, R$ }{

(1) Perform LMMSE channel estimation to obtain  $\hat{\bH}^{(r)}$

(2) Perform distributed EP-based MIMO detection to obtain  $\hat{\bX}^{(\mathrm{d}, r)}$

(3) Data feedback to channel estimator
}
\end{algorithm}
\subsection{LMMSE Estimator}\label{SEC:SE}
We consider adopting the classical LMMSE estimator in the channel estimation stage at each AP. 
To facilitate the representation of the channel estimation problem, we consider the matrix vectorization to (\ref{eqLMMSE}) and rewrite it as
\begin{equation}\label{eq_vec_ce}
\by^{\rmp}_{l} = \bA^{\rmp} \bar{\bh}_{l} + \bn^{\rmp}_{l},
\end{equation}
where $\bA^{\rmp} = (\bX^{\rmp})^{T} \otimes \bI_{N} \in \mathbb{C}^{\tau_{p}N \times KN}$,  $\by^{\rmp}_{l}=\mathrm{vec}(\bY^{\rmp}_{l}) \in \mathbb{C}^{\tau_{p} N \times 1}$, $\bar{\bh}_{l} = \mathrm{vec}(\bH_l) \in \mathbb{C}^{N K \times 1} $,  and $ \bn^{\rmp}_{l} = \mathrm{vec}(\mathbf{N}^{\rmp}_{l})\in \mathbb{C}^{\tau_{p}N \times 1}$.  We denote $\otimes$  as the matrix Kronecker product and $\mathrm{vec}(\cdot)$  as the vectorization operation. In the pilot-only based channel estimation stage, the LMMSE estimate of $\bar{\bh}_{l}$ is given by
\begin{equation}\label{LMMSE}
 \hat{\bar{\bh}}^\rmp_{l} = \bR_{\bar{\bh}_{l} \bar{\bh}_{l}}(\bA^{\rmp})^{H}(\bA^{\rmp} \bR_{\bar{\bh}_{l} \bar{\bh}_{l}} (\bA^{\rmp})^{H}+\sigma^{2}\bI_{\tau_{p} N})^{-1}\by_{l}^{\rmp},
\end{equation}
where $\bR_{\bar{\bh}_{l} \bar{\bh}_l}$ is the channel covariance matrix and depends on the spatial correlation matrix $\bR_{kl}$. Based on the property of the LMMSE estimator, $\hat{\bar{\bh}}_l^{\rmp}$ is a Gaussian random vector, and the channel estimation error vector $\Delta{\bh^{\rmp}_l}=\hat{\bar{\bh}}^{\rmp}_l-\bh_l$ is also a Gaussian random vector with zero-mean and  covariance matrix $ \bR_{\Delta{\bh_l^\rmp}}$  given by
\begin{equation}\label{LMMSE_error}
 \bR_{\Delta{\bh_l^\rmp}} = \bR_{\bar{\bh}_{l} \bar{\bh}_{l}}-\bR_{\bar{\bh}_{l} \bar{\bh}_{l}}(\bA^{\rmp})^{H}(\bA^{\rmp} \bR_{\bar{\bh}_{l} \bar{\bh}_{l}} (\bA^{\rmp})^{H}+\sigma^{2}\bI_{\tau_{p}N})^{-1}\bA^{ \rmp}\bR_{\bar{\bh}_{l} \bar{\bh}_{l}}.
\end{equation}
\subsection{Signal Detection with Channel Estimation Errors}\label{SD_CE}
After performing channel estimation, the estimated channel is used to perform data detection at each AP. In the data transmission stage, the received data signal vector $\by^{\rmd}[n]$ corresponding to the $n$-th data in each coherence time can be expressed by
$\by^{\rmd}[n]=\bH\bx^{\rmd}[n]+\bn^{\rmd}[n]$, where $\bn^{\rmd}[n] \sim \mathcal{N}_{\mathbb{C}}(0,\sigma^{2}\bI_N) $  is the AWGN vector\footnote{{\blc As each AP has similar signal models, we remove the index $l$ for the AP and use $\by^{\rmd}[n]$, $\bH$, $\bx^{\rmd}[n]$, and $\bn^{\rmd}[n]$ to denote $\by_{l}^{\rmd}[n]$, $\bH_{l}$, $\bx_{l}^{\rmd}[n]$, and $\bn_{l}^{\rmd}[n]$, respectively. To simplify the notation, we have the same operation for the following symbols.}}. We can rewrite it in a matrix from as $\bY^{\rmd}=\bH\bX^{\rmd}+\bN^{\rmd}$, where $\bN^{\rmd} = \left[\bn^{\rmd}[1],\ldots, \bn^{\rmd}[\tau_d] \right]\in \bbC^{N \times \tau_d} $ is the AWGN matrix in the data transmission stage.
Denote the estimated channel $\hat{\bH} \in \mathbb{C}^{N\times K}$ as $\hat{\bH} = \bH + \Delta\bH$,
where $\Delta\bH$ is the channel estimation error. The signal detection problem can be formulated as
\begin{align}\label{eqSD_CSI}
  \by^{\rmd}[n]  =\bH\bx^{\rmd}[n]+\bn^{\rmd}[n] = \hat{\bH}\bx^{\rmd}[n] + \hat{\bn}^{\rmd}[n],
\end{align}
where $\hat{\bn}^{\rmd}[n] = \bn^{\rmd}[n] - \Delta\bH\bx^{\rmd}[n]$ is the equivalent noise in the signal detector and the statistical characterization of $\hat{\bn}^{\rmd}[n]$  is given in  Appendix \ref{derivation21}.
\subsection{Data-Aided Channel Estimation}\label{Data_aided}
To further improve the BER performance, a data-aided channel estimation approach is adopted in the  channel estimation stage. First, conventional pilot-only based channel estimation is performed and the transmitted symbols are detected. Then, the detected symbols are fed back to the channel estimator as additional pilot symbols to refine the channel estimation.  In the channel training stage, pilot matrix $\bX^{\rmp}$ is transmitted. The received signal matrix corresponding to the pilot matrix $\bY^{\rmp}\in\mathbb{C}^{N\times \tau_p}$  is expressed as $\bY^{\rmp}=\bH\bX^{\rmp}+\bN^{\rmp}$,
where $\bN^{\rmp} = \left[\bn^{\rmp}[1],\ldots,\mathbf{n}^{\rmp}[\tau_p]\right]  \in \mathbb{C}^{N\times \tau_p}$  is the AWGN matrix and each column $ \mathbf{n}^{\rmp}[n] \sim \mathcal{N}_{\mathbb{C}}(0,\sigma^{2}\mathbf{I}_{N}) $ for $n = 1,\ldots, \tau_p $. 
The estimated data matrix $\hat{\mathbf{X}}^{\rmd}$ can be expressed as $\hat{\bX}^{\rmd} = \bX^{\rmd} + \bE^{\rmd}$,
where $\mathbf{E}^{\rmd}$ is the signal detection error matrix. In the data-aided channel estimation stage, the estimated $\hat{\bX}^{\rmd}$ are fed back to the channel estimator as additional pilots. Then, the received signal matrix $\bY^{\rmd}$ corresponding to $\hat{\bX}^{\rmd}$ can be expressed as
\begin{align}\label{eqCE_data1}
\bY^{\rmd}  = \bH\bX^{\rmd}+\bN^{\rmd} =  \bH \hat{\bX}^{\rmd}+\hat{\bN}^{\rmd},
\end{align}
where $\hat{\bN}^{\rmd} = \bN^{\rmd} - \bH\bE^{\rmd}$  is the equivalent noise for the data-aided pilot  $\hat{\bX}^{\rmd}$. The statistical information of the $n$-th column of $\hat{\bN}^{\rmd}$ is $\hat{\mathbf{n}}^{\rmd}[n]\sim \mathcal{N}_{\bbC}(\mathbf{0},\hat{\mathbf{V}}^{\rmd}[n])$ for $n=1,\ldots,\tau_d $, where $\hat{\mathbf{V}}^{\rmd}[n]$  will be utilized for data-aided channel estimation. We denote $\bY = [\bY^{\rmp}, \bY^{\rmd}]$ as the received signal matrix corresponding to the overall transmitted signal. By concatenating the received pilot and data signals, we have
\begin{align}\label{eqCE_data}
\bY  = \left[\bY^{\rmp} \ \bY^{\rmd}\right]  = \bH\bX+\bN,
\end{align}
where $\bX = [\bX^{\rmp}, \hat{\bX}^{\rmd}]$ and  $\bN = [\bN^{\rmp}, \hat{\bN}^{\rmd}]$
can be interpreted as the equivalent pilot signal and noise in the data-aided channel estimation stage, respectively. The statistical characteristic for the equivalent noise and  corresponding LMMSE estimator are derived in Appendix \ref{derivation22}.
\section{Performance Analysis}\label{SE:performance}
In this section, we provide an asymptotic performance analysis for the proposed distributed EP detector. We consider $L,K \rightarrow \infty$ and fix
\begin{equation}\label{eqratio}
  \alpha_{l} = \frac{K}{N}, \alpha = \frac{K}{LN}.
\end{equation}
After proposing a state evolution analysis framework, we will characterize the fixed points of the proposed detector.
\subsection{State Evolution Analysis}\label{SEC:SE}
We first provide a state evolution analysis framework to predict the asymptotic performance of the distributed EP detector in the large-system limit\footnote{The large-system limit means that the cell-free massive MIMO system with $L,K \rightarrow \infty$.}.
\begin{proposition}\label{proposition1}
In the large-system limit, the asymptotic behavior (such as MSE and BER) of  \textbf{Algorithm 1} can be described by the following equations:
\begin{subequations} \label{state_evolution}
   \begin{align}
     v_{\mathrm{A},l}^{\mathrm{ext},t} & = \bigg( \frac{1}{K_{l}} \sum_{k=1}^{K_l}\frac{1}{\sigma^{-2}\tau_{k,l}+\lambda_{l}^{(t-1)}}-\lambda_{l}^{(t-1)}\bigg)^{-1} \\
   v_{\mathrm{A}}^{\mathrm{ext},t} & = \bigg( \sum_{l=1}^{L} \frac{1}{ v_{\mathrm{A},l}^{\mathrm{ext}, t}}\bigg)^{-1} \\
   \lambda_{l}^{(t)} & = \frac{1}{\mathrm{MSE}(v_{\mathrm{A}}^{\mathrm{ext},t})}-\frac{1}{v_{\mathrm{A},l}^{\mathrm{ext},t}}.
   \end{align}
\end{subequations}
\hfill\ensuremath{\blacksquare}
\end{proposition}
The function $\mathrm{MSE}(\cdot)$ is given by
\begin{equation}\label{eqMSE}
\mathrm{MSE}(v_{\mathrm{A},l}^{\mathrm{ext},t}) \triangleq \mathtt{E}\bigg\{| x-\mathtt{E}\{x|x_{\mathrm{A},l}^{\mathrm{ext},t},v_{\mathrm{A},l}^{\mathrm{ext},t}|^{2}\bigg\}
\end{equation}
where the expectation is with respect to $x$. $\tau_{k,l}$ are the eigenvalues of $\bH^{H}_{l}\bH_{l}$.

The state equations illustrated in Proposition \ref{proposition1} can be proved rigorously in a similar way as \cite{Takeuchi}. To better understand the state evolution analysis, we give an  intuitive derivation here. In \textbf{Algorithm 1}, we have explicit expressions for computing the mean and variance for each module. By substituting (\ref{eqvarA}) into (\ref{eqextvar}), we have the following expression
\begin{equation}\label{eqextvarasy}
v_{\rmA,l}^{\mathrm{ext},t} = \frac{1}{\frac{1}{K}\mathrm{tr}(\sigma^{-2} \bH^{H}_{l}\bH_{l}+\lambda_{l}^{(t-1)}\bI)}-\lambda_{l}^{(t-1)}.
\end{equation}

The asymptotic expressions for $v_{\mathrm{A}}^{\mathrm{ext},t}$ and $\lambda_{l}^{(t)}$ can then be derived from (\ref{eqvarmrc}) and (\ref{eqvarest}).
Finally, the asymptotic MSE can be interpreted as the MSE of the decoupled scalar AWGN channels (\ref{eqAWGN}) and is related to the distribution of the transmitted signal $\bx$.  Next, we will give a specific example for \textbf{Proposition 1}.

\emph{ Example~1:} If the channel matrix $\bH_{l}\,(l=1,2,\ldots,L)$  is an i.i.d. Gaussian distributed matrix, then $v_{\rmA,l}^{\mathrm{ext}}$  will converge to the following asymptotic expression by adopting the result of the $\mathcal{R}$-transform\footnote{{\blc The concept of $\mathcal{R}$-transform can be found in \cite{Random_matrix}, which enables the characterization of the asymptotic spectrum of a sum of
suitable matrices from their individual asymptotic spectra. }}  of the average empirical eigenvalue distribution given by,

\begin{equation}\label{eqconverge}
    v_{\mathrm{A},l}^{\mathrm{ext},t} = \frac{\alpha_{l}\sigma^{2}+(\alpha_{l}-1)\lambda_{l}^{(t-1)}}{2}
    +\frac{\sqrt{\bigg(\alpha_{l}\sigma^{2}+(\alpha_{l}-1)\lambda_{l}^{(t-1)}\bigg)^{2}+4\alpha_{l}\sigma^{2}\lambda_{l}^{(t-1)}}}{2}.
\end{equation}
Furthermore, if the data symbol is drawn from a quadrature phase-shift keying (QPSK) constellation, the $\mathrm{MSE}$ is expressed by

\begin{equation}\label{mse_x_explicit expression}
  \mathrm{MSE}(v_{\mathrm{A},l}^{\mathrm{ext},t}) = 1-\int \rmD z\tanh(\frac{1}{v_{\mathrm{A},l}^{\mathrm{ext},t}}+\sqrt{\frac{1}{v_{\mathrm{A},l}^{\mathrm{ext},t}}}z).
\end{equation}
Furthermore, the BER w.r.t. $\bx$ can  be evaluated through the equivalent AWGN channel (\ref{eqAWGN}) with an equivalent $\mathrm{SNR} = 1/v_{\mathrm{A},l}^{\mathrm{ext},T}$ and is given by
\begin{equation}\label{SER_QPSK}
  \mathrm{BER}= 2Q(\sqrt{\frac{1}{v_{\mathrm{A},l}^{\mathrm{ext},T}}})-[Q(\sqrt{\frac{1}{v_{\mathrm{A},l}^{\mathrm{ext},T}}})]^{2},
\end{equation}
where $Q(x)= \int_{x}^{\infty}\rmD z$ is the $Q$-function. In fact, the MSE and BER are determined given the knowledge of the  AWGN channel (\ref{eqAWGN}) with $\mathrm{SNR} = 1/v_{\mathrm{A},l}^{\mathrm{ext},T}$, which is known as the decoupling principle. Thus, if the data symbol is drawn from other $M$-QAM constellations, the corresponding BER can be easily obtained using a closed-form BER expression \cite{Proakis2007}.
\subsection{Fixed Point Characterization}\label{SEC:fixed}
The fixed points of the distributed EP detection algorithm are the stationary points of a relaxed version of the Kullback-Leibler (KL) divergence to minimize the posterior probability (\ref{eq4}), which is given by
\begin{equation}\label{eq_KL}
b(\bx)=\arg\min\limits_{b(\bx)\in\cF}D\left[ b(\bx) \| \mathtt{P}(\bx|\by,\bH)\right].
\end{equation}
The KL divergence between two distributions is defined as,
\begin{equation}\label{KL_Definition}
	{D}\left[ {p\left( x \right)\left\| {q\left( x \right)} \right.} \right]=\int {p\left( x \right)\log \frac{{p\left( x \right)}}{{q\left( x \right)}}} {\rmd}x +\int{\left( {q\left( x \right)-p\left( x \right)} \right)} {\rmd}x.
\end{equation}
Generally, it is difficult to approximate the complex posterior distribution (\ref{eq4}) by a tractable approach. The {\em moment matching} method is utilized to exploit the  relaxations of the KL divergence minimization problem \cite{GEC_Rangan}. The process for solving the KL divergence minimization problem is equivalent to optimizing the Bethe free energy subject to the moment-matching constraint. That is,
\begin{subequations}\label{BFE_Min}
\begin{align}
			&\min\limits_{b,b_1,\ldots,b_L}\max\limits_{q}~J\left(b,b_1,\ldots,b_L,q\right) \label{Bethe_Free_Energy}  \\
			&~~~~~\text{s.t.}~~~~\mathtt{E}_{b_l}(\bx)=\mathtt{E}_{q}(\bx),\displaybreak[0]\label{BFE_Min_Const1}\\
			&~~~~~~~~~~~~\frac{1}{K}\sum\limits_{k\in\mathcal{K}}\mathtt{E}_{b_l}(|x_k|^2)= \frac{1}{K}\sum\limits_{k\in\mathcal{K}}\mathtt{E}_{q}(|x_k|^2),\label{BFE_Min_Const2}
\end{align}
\end{subequations}
where $J\left(b,b_1,\ldots,b_L,q\right) =\mathrm{KL}\left(b\|e^{\log f(\bx)}\right)+\sum\limits_{l=1}^{L} \mathrm{KL}\left(b_l\|e^{\log f_L(\bx)}\right)+H(q)$ and
$H(\cdot)$ is the differential entropy. In addition, $b(\bx)$, $\{b_c(\bx)\}_{l=1}^{L}$, and $q(\bx)$ are given by
\begin{subequations}\label{stat_points}
\begin{align}			&b(\bx):=\frac{1}{Z_0\left(\bx_{\rmA}^{\mathrm{ext}}\right)}p(\bx)\exp\left[-\|\bx-\bx_{\rmA}^{\mathrm{ext}}\|^2/v_{\rmA}^{\mathrm{ext}}\right],\displaybreak[0]\label{Unbiased_Est_x}\\
&b_l(\bx):=\frac{1}{Z_c\left(\bx_{\rmB,l}^{\mathrm{ext}}\right)}\exp\left[-\frac{1}{\sigma^{2}}\|\by_l-\bh_l\bx\|^{2}-\|\bx-\bx_{\rmB,l}^{\mathrm{ext}}\|^2/v_{\rmB,l}^{\mathrm{ext}}\right],\displaybreak[0]\\&q(\bx):=\frac{1}{Z_q\left(\hat{\bx}\right)}\exp\left[-\|\bx-\hat{\bx}\|^2/v\right],
\end{align}
\end{subequations}
respectively, where $Z_0\left(\bx_{\rmA}^{\mathrm{ext}}\right)$, $\{Z_c\left(\bx_{\rmB,l}^{\mathrm{ext}}\right)\}_{c=1}^{C}$, and $Z_q\left(\hat{\bx}\right)$ are the normalization factors  corresponding to their density functions. 
Based on the above-mentioned optimization process, we have
\begin{subequations}\label{Moment_Matching}
	\begin{align}
	&\hat{\bx}=\mathtt{E}_{b}(\bx)=\mathtt{E}_{b_1}(\bx)=\cdots=\mathtt{E}_{b_C}(\bx)= \mathtt{E}_{q}(\bx),\\
	&\begin{aligned}
	v&=\frac{1}{K}\sum\limits_{k\in\mathcal{K}}\mathtt{E}_{b}(|x_k|^2)=\frac{1}{K}\sum\limits_{k\in\mathcal{K}}\mathtt{E}_{b_1}(|x_k|^2)=\cdots
	&=\frac{1}{K}\sum\limits_{k\in\mathcal{K}}\mathtt{E}_{b_L}(|x_k|^2)= \frac{1}{K}\sum\limits_{k\in\mathcal{K}}\mathtt{E}_{q}(|x_k|^2),	
	\end{aligned}
	\end{align}
\end{subequations}
for any fixed points $v$ and $\hat{\bx}$ in \textbf{Algorithm 1}. {\blc This fixed point characterization is helpful to perform the convergence analysis and understand the algorithm from the optimization perspective.}
%
\section{Simulation Results}\label{simulation}
In this section, we will provide extensive simulation results to demonstrate the excellent performance of the proposed distributed EP detector for cell-free massive MIMO. After showing the simulation results of the distributed EP detector in conventional and scalable cell-free massive MIMO systems, we will verify the accuracy of the proposed analytical framework. Furthermore, the performance of the distributed EP-based ICD scheme is also investigated. The codes for the simulation part is available at https://github.com/hehengtao/DeEP-cell-free-mMIMO.
\subsection{Conventional Cell-Free Massive MIMO}\label{sec:simulation1}
\begin{figure}
\begin{minipage}{3in}
  \centerline{\includegraphics[width=3.0in]{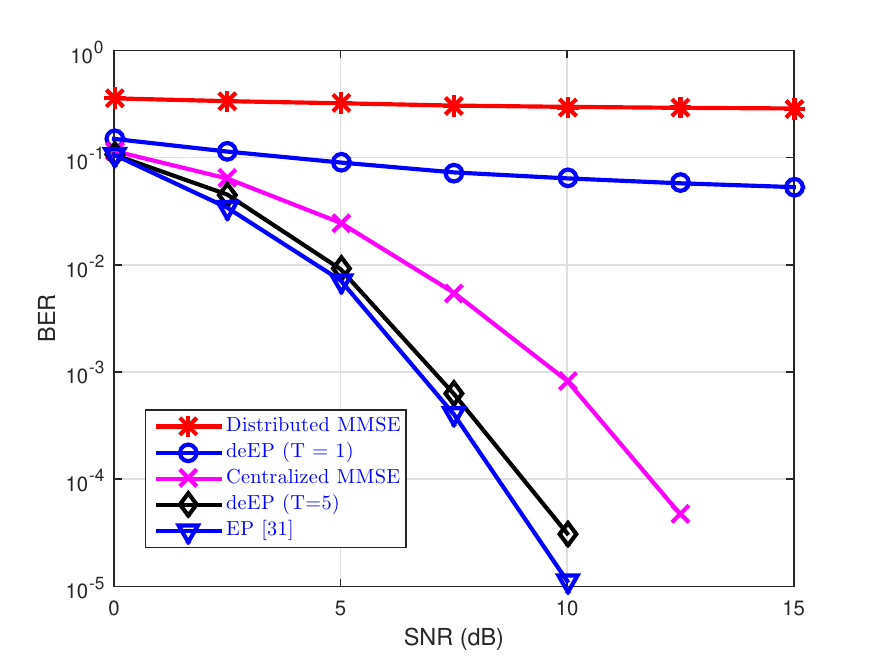}}
  \centerline{(a) QPSK}
\end{minipage}
\hfill
\begin{minipage}{3in}
  \centerline{\includegraphics[width=3.0in]{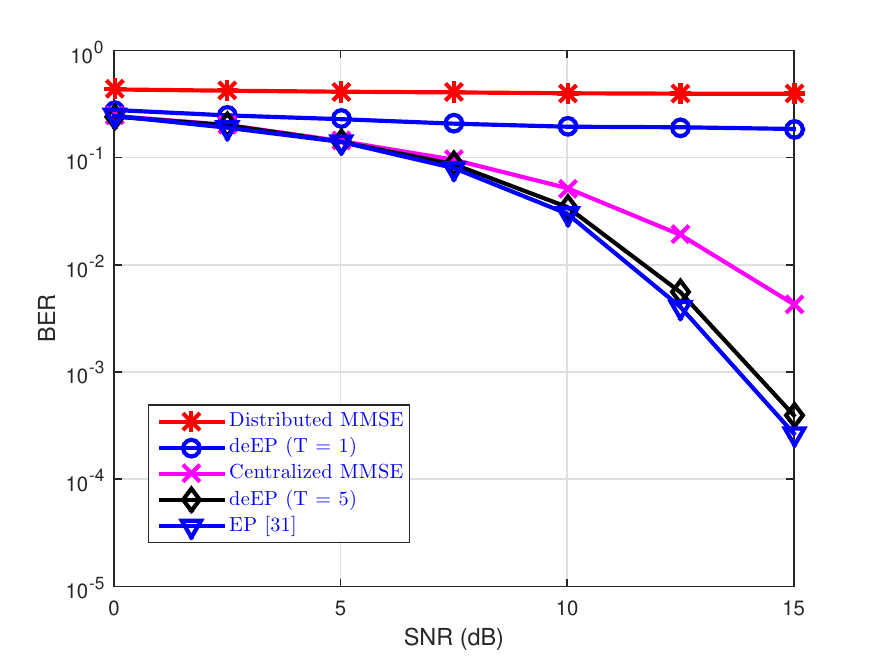}}
  \centerline{(b) 16-QAM}
\end{minipage}
\caption{.~~BER performance comparisons of different detectors in the conventional cell-free massive MIMO system.}
\label{Fig:rayleigh}
\end{figure}

In this subsection, we consider the BER performance of the proposed detector in the conventional cell-free massive MIMO where each AP serves all users. We assume that perfect CSI can be obtained at each AP. The simulation parameters are set as  $N=8$, $K=32$, and $L=8$. {\blc Fig.\,\ref{Fig:rayleigh} compares the achievable BER of the proposed distributed EP, centralized EP detectors \cite{EP_detector}, and other baselines \cite{cell_free_distributed_receiver} using QPSK and $16$-QAM modulation symbols.} The results are obtained by Monte Carlo simulations with $10,000$ independent channel realizations. We denote ``deEP" as the distributed EP detector proposed in Section \ref{SEC:EP}. It can be observed that the proposed distributed EP detector outperforms the distributed MMSE  detector with only one EP iteration. This is because  deEP incorporates the prior information of the symbol $\bx$ into the posterior mean estimator for the equivalent AWGN channel (\ref{eqAWGNl}). Furthermore, it outperforms the centralized MMSE detector with $T=5$ iterations for different modulation symbols. {\blc Compared with the centralized EP detectors \cite{EP_detector}, the distributed EP detector achieves comparable BER performance.}
\subsection{Accuracy of the Analytical Results}\label{sec:simulation2}
\begin{figure}
\begin{minipage}{3in}
  \centerline{\includegraphics[width=3.0in]{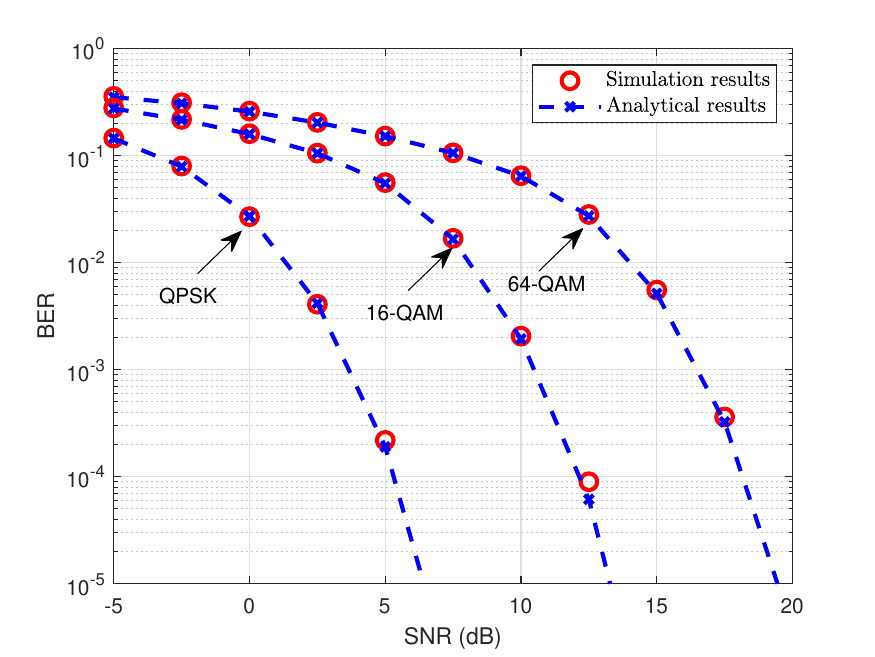}}
  \centerline{(a) {\blc $N=16$, $K=64$, $L=16$}}
\end{minipage}
\hfill
\begin{minipage}{3in}
  \centerline{\includegraphics[width=3.0in]{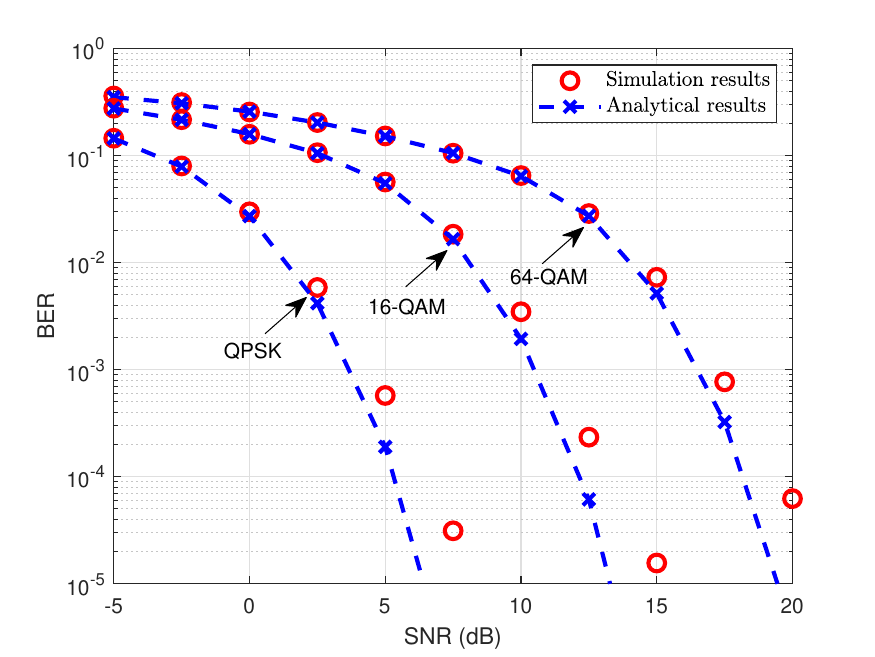}}
  \centerline{(b) {\blc $N=8$, $K=8$, $L=4$}}
\end{minipage}
\caption{.~~BER performance comparisons of the analytical and simulation results for the distributed EP detector with different modulation schemes.}
\label{Fig:analytical}
\end{figure}

We next verify the accuracy of the analytical framework in Fig.\,\ref{Fig:analytical} with {\blc different system settings and modulation schemes.} As shown in the figure, the BERs of the proposed detector match well with the derived analytical results, which demonstrates the accuracy of the analytical framework developed in Section \ref{SEC:SE}. {\blc Although the analytical framework is derived in the large-system limit, it still can predict the BER performance in small-sized MIMO systems illustrated in Fig.\,\ref{Fig:analytical}(b).} Therefore, instead of performing time-consuming Monte Carlo simulations to obtain the corresponding performance metrics, we can predict the theoretical behavior by the state evolution equations. Furthermore, the analytical framework can be further utilized to optimize the system design.
\subsection{Scalable Cell-Free Massive MIMO}\label{sec:simulation3}
\begin{figure}
\begin{minipage}{3in}
  \centerline{\includegraphics[width=3.0in]{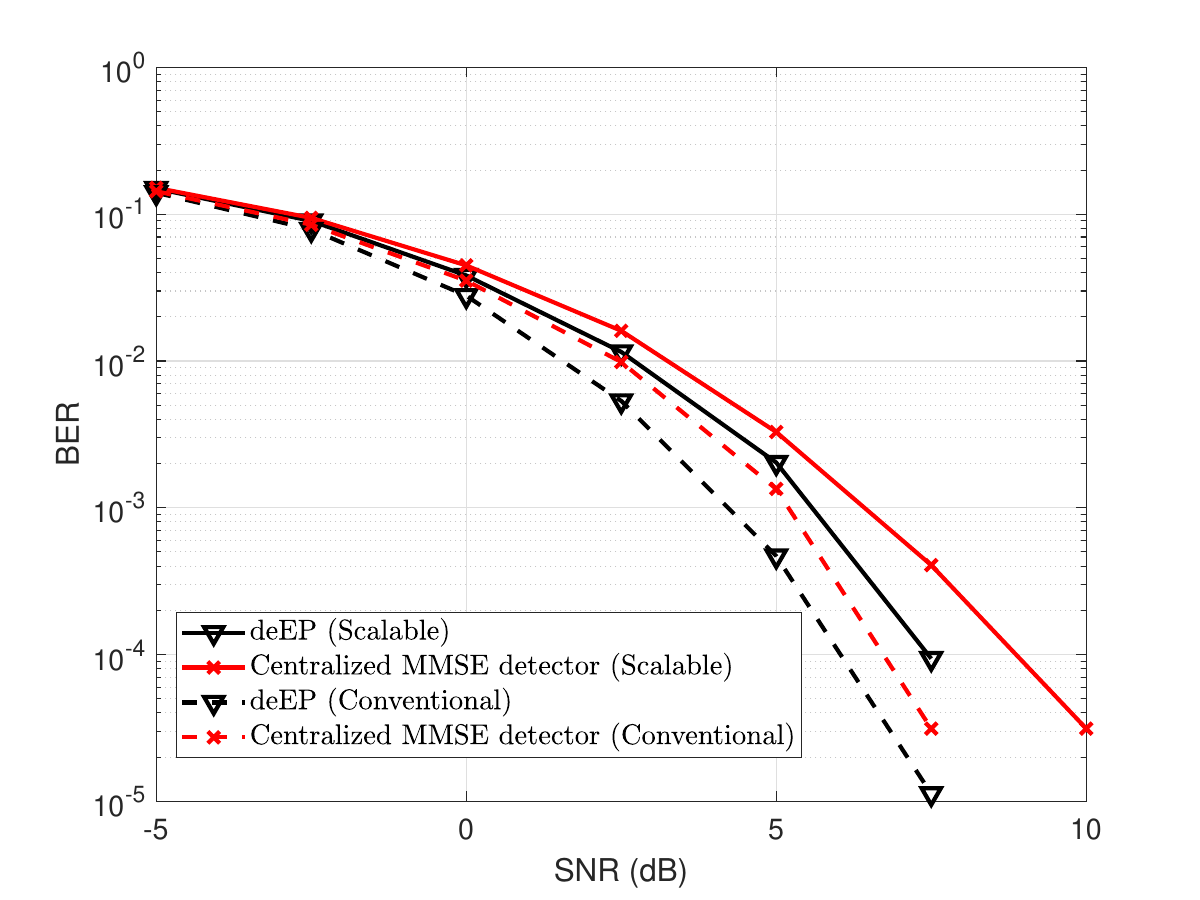}}
  \centerline{(a) QPSK}
\end{minipage}
\hfill
\begin{minipage}{3in}
  \centerline{\includegraphics[width=3.0in]{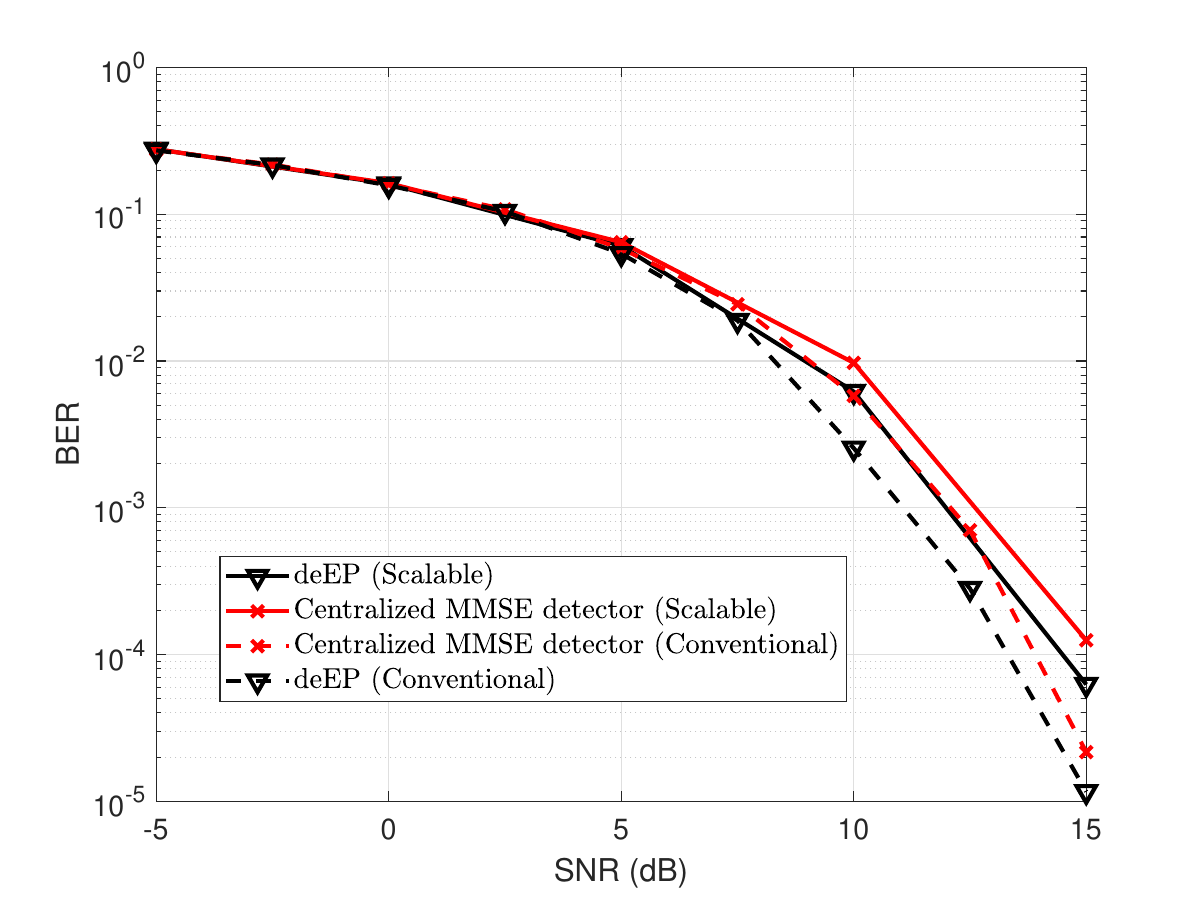}}
  \centerline{(b) 16-QAM}
\end{minipage}
\caption{.~~BER performance comparisons of different detectors in the scalable cell-free massive MIMO system.}
\label{Fig:scalable}
\end{figure}

As the \emph{user-centric} approach is more attractive for cell-free massive MIMO, we investigate the distributed EP detector with DCC framework. The user first appoints a master AP according to the large-scale fading factor and assigned a pilot $\bp_{\tau}$  by the appointed AP. Then, other neighboring APs determine whether they serve the accessing user according to the assigned pilot $\bp_{\tau}$. Finally, the cluster for the $k$-th user is constructed. Fig.\,\ref{Fig:scalable} shows that the distributed EP detector outperforms both the centralized and distributed MMSE detectors. Furthermore, the performance loss is limited when compared to the conventional cell-free massive MIMO system while the computational complexity is significantly decreased from $\mathcal{O}(KN^{2})$ to $O(|\mathcal{D}_{l}|N^{2})$.
\subsection{Performance of ICD with 3GPP Urban Channel}\label{sec:simulation4}
In  Sections \ref{sec:simulation1} to \ref{sec:simulation4}, all detectors are investigated assuming perfect CSI. In this section, we consider a distributed EP-based ICD scheme for cell-free massive  MIMO  with the 3GPP Urban Channel \cite {3GPP} and imperfect CSI. We assume that $N=8$, $K=8$, and $L=4$. The APs are assumed to be deployed in a $1\times1$ $\mathrm{km}^2$ area located in urban environments to match with the system settings in \cite{cell_free_distributed_receiver}. The large-scale fading factor is given by
\begin{equation}
\beta_{kl} \,  [\textrm{dB}] = -30.5 - 36.7 \log_{10}\left( \frac{d_{kl}}{1\,\textrm{m}} \right)  + g_{kl},
\end{equation}
where $d_{kl}$ is the distance between the $k$-th user and $l$-th AP and $g_{kl} \sim \mathcal{N}(0,4^2)$ represents shadow fading. The shadowing terms from an AP to different users are correlated as
\begin{equation} \label{eq:shadowing-decorrelation}
\mathbb{E} \{ g_{kl} g_{ij} \} =
\begin{cases}
4^2 2^{-\delta_{ki}/9\,\textrm{m}} & l = j \\
0 & l \neq j,
\end{cases}
\end{equation}
where $\delta_{ki}$ is the distance between the $k$-th user and $i$-th user. The second case in \eqref{eq:shadowing-decorrelation} on the correlation of shadowing terms related to two different APs can be ignored. Furthermore, the multi-antenna APs are equipped with half-wavelength-spaced uniform linear arrays and the spatial correlation is generated using the Gaussian local scattering model with a $15^\circ$ angular standard deviation. The transmit power $p_k$ for each user is  $p_k = 100$\,mW and the bandwidth is $20$\,MHz.
\begin{figure}
\begin{minipage}{3in}
  \centerline{\includegraphics[width=3.0in]{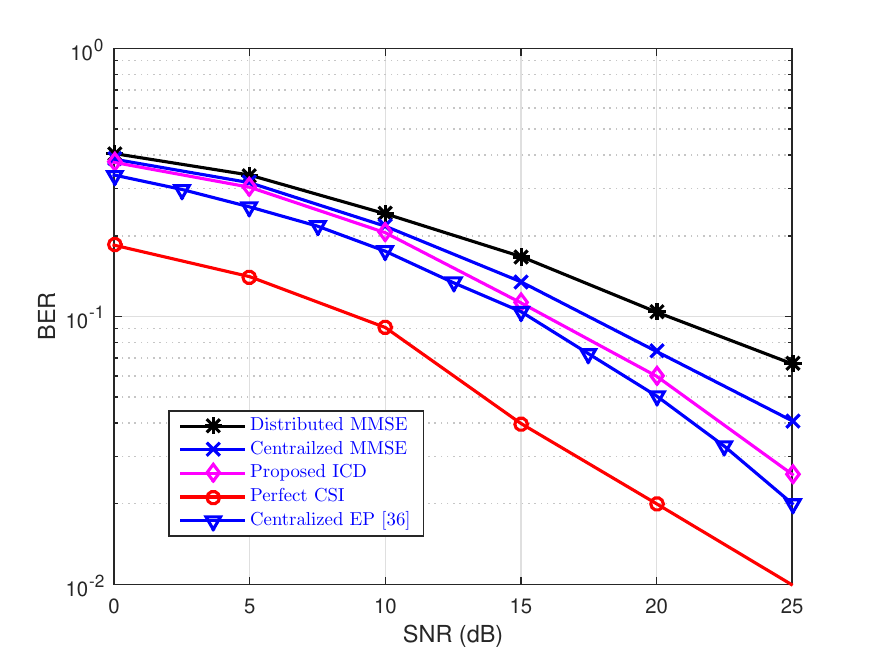}}
  \centerline{(a) QPSK}
\end{minipage}
\hfill
\begin{minipage}{3in}
  \centerline{\includegraphics[width=3.0in]{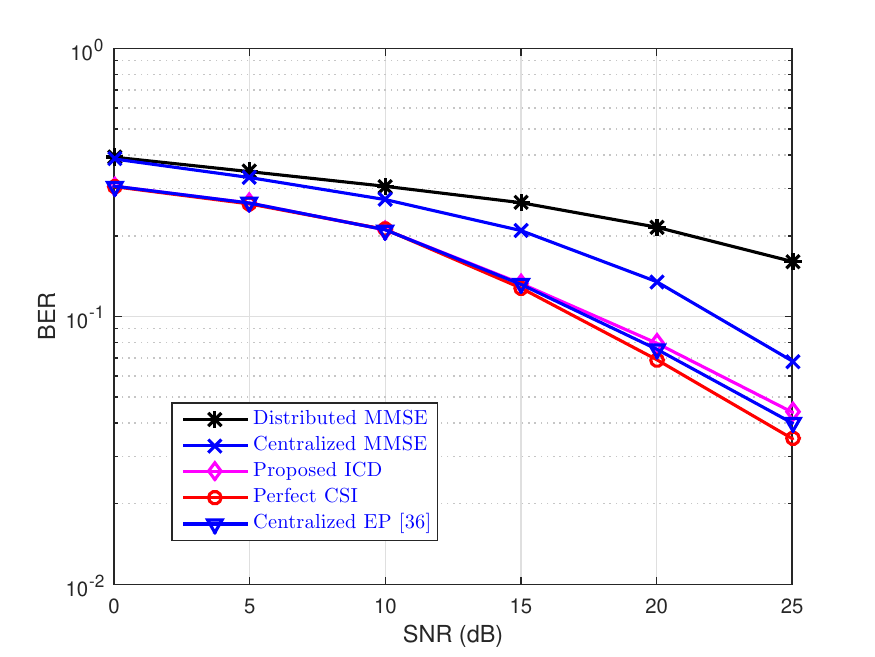}}
  \centerline{(b) 16-QAM}
\end{minipage}
\caption{.~~BER performance comparisons of different detectors in the cell-free massive MIMO system with 3GPP Urban Channel model and imperfect CSI.}
\label{Fig:large}
\end{figure}
Fig.\,\ref{Fig:large} compares the achievable BER of the proposed distributed EP detector with those of other detectors in the 3GPP Urban Channel and imperfect CSI. The channel is estimated with the LMMSE channel estimator. As can be observed from the figure, the proposed ICD scheme outperforms the centralized and distributed MMSE detectors, and shows a similar performance as the distributed EP detector with perfect CSI and $16$-QAM symbols.

\subsection{Non-Orthogonal Pilots}\label{sec:simulation5}
\begin{figure}
\begin{minipage}{3in}
  \centerline{\includegraphics[width=3.0in]{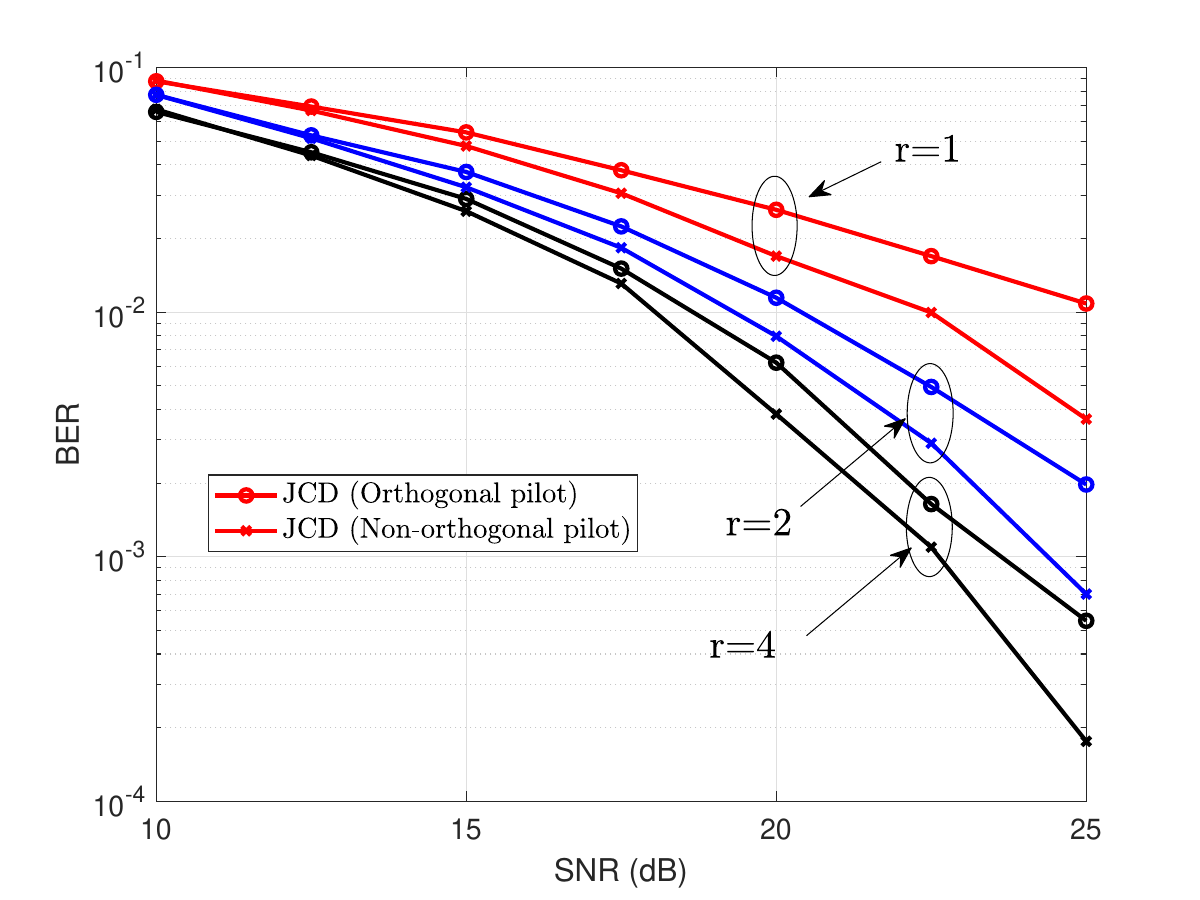}}
  \centerline{(a) QPSK}
\end{minipage}
\hfill
\begin{minipage}{3in}
  \centerline{\includegraphics[width=3.0in]{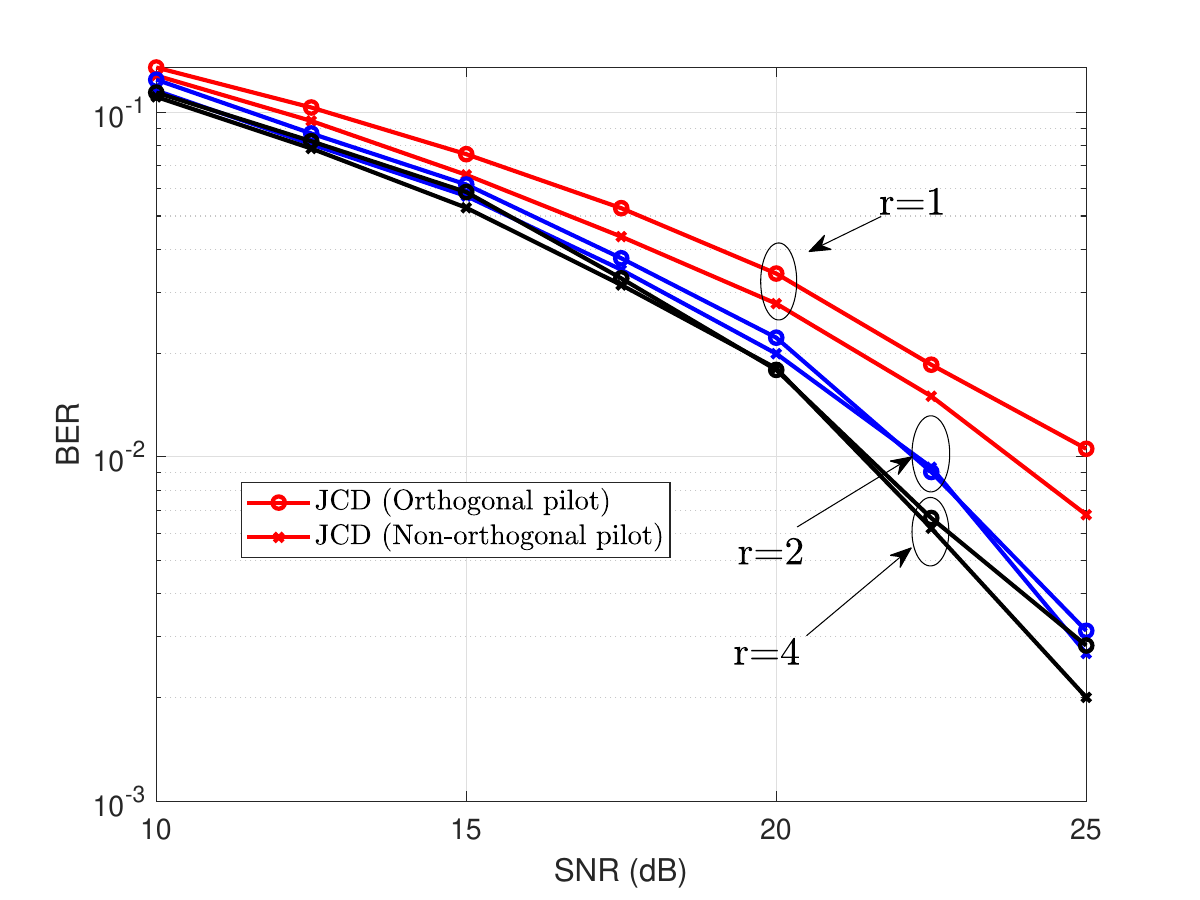}}
  \centerline{(b) 16-QAM}
\end{minipage}
\caption{.~~BERs performance comparisons of the distributed EP-based ICD scheme with orthogonal and non-orthogonal pilots.}
\label{Fig:ICD}
\end{figure}

In this section, we consider the performance of ICD with different pilots in the channel training stage. In particular, the orthogonal pilot matrix $\mathbf{X}^{\rmp} \in \mathbb{C}^{ N\times \tau_p}$ is chosen by selecting $\tau_p$ columns from the discrete Fourier transformation (DFT) matrix $\bF \in \mathbb{C}^{\tau_p\times \tau_p}$. For a non-orthogonal pilot, each element of the pilot matrix is drawn from $64$-QAM.  {\blc Furthermore, we consider $\tau_{p} = K$ and $\tau_{d} = 16K$ for the two pilots.} Fig.\,\ref{Fig:ICD} shows the BERs of the distributed detectors in the ICD architecture, where $r=1$ indicates no data feedback to the channel estimator while $r=2$ refers to the scenario where the detected data are fed back to channel estimator once. As can be observed from the figure, the data feedback can improve the system performance significantly. Specifically, if we target for a BER = $10^{-2}$ with QPSK symbols and non-orthogonal pilots, data feedback ($r=2$) can bring  $3.3$ dB  and data feedback ($r=4$) can bring $4.5$ dB performance gain, respectively. Furthermore, the performance gain will be enlarged if orthogonal DFT pilots are used. Interestingly, using DFT pilots has a similar performance as that of non-orthogonal pilots. This demonstrates that the ICD scheme is an efficient way to enable non-orthogonal pilots for reducing the pilot overhead in cell-free massive MIMO systems.
\section{Conclusion}\label{con}
In this paper, we proposed a distributed EP detector for cell-free massive MIMO. It was shown that the proposed detector achieves better performance than  linear detectors for both conventional and scalable cell-free massive MIMO networks. Compared to other distributed detectors, it achieves a better BER performance with an increase of the computational overhead at the CPU. A distributed ICD was then proposed for cell-free massive MIMO systems to handle imperfect CSI. An analytical framework was also provided to characterize the asymptotic performance of the proposed detector in a large system setting. Simulation results  demonstrated  that the proposed method outperforms existing distributed detectors for cell-free massive MIMO in terms of BER performance. Furthermore, the proposed ICD architecture significantly improves the system performance and enables non-orthogonal pilots to reduce the pilot overhead. For future research, it is  interesting to extend the distributed EP algorithm to precoding and channel estimation for cell-free massive MIMO systems.

\appendices

\section{Message-Passing Derivation of Algorithm 1}
\subsection{Derivation of Distributed EP Detector}\label{derivation}
{\blc In this appendix, we present the derivation of \textbf{Algorithm 1} from the message-passing perspective\footnote{{\blc We also can derive \textbf{Algorithm 1} by directly utilizing the EP principle in \cite{EP1}, which is an extension of belief propagation and obtained by imposing an exponential family constraint on  messages. Both BP and EP can be derived from the generic message passing rules introduced in Section \ref{Sec:message_passing}.}}.} By applying the above message-passing rules to the factor graph in Fig.\,\ref{Fig:factor_graph}, we can obtain \textbf{Algorithm 1}. First, we give the factor graph for cell-free massive MIMO detection. As illustrated in Fig.\,\ref{Fig:factor_graph}, we have factor graphs for describing the centralized processing model (\ref{eq4}) and distributed processing model (\ref{eqdistributed}). Specifically, the factor graph contains $L+1$ factor nodes $\{f,f_1,\ldots,f_L\}$ and one variable node $\bx$ for the distributed processing model (\ref{eqdistributed}). The factor $f$ represents the prior information of $\bx$ and  $f_l$ denotes $l$-th AP. After constructing the factor graph for (\ref{eqdistributed}),  we approximate the posterior distribution by computing and passing messages among the nodes in the factor graph with an iterative manner and aforementioned rules. We initialize the messages from the variable node $\bx$ to factors $\{f_0,f_1,\ldots,f_L\}$ as $\mu_{\bx \rightarrow f_{l}}(\bx)= \mathcal{N}_{\mathbb{C}}(\bx; \boldsymbol{\gamma}_{l}^{(0)}/\lambda_{l}^{(0)}, \lambda_{l}\bI) (l=1,2,...,L)$. {\blc Then, according to Factor-to-variable messages, we have the message  $ \mu_{f_l\rightarrow \bx}(\bx)  = b_{f_{l}}(\bx)/\mu_{\bx \rightarrow f_{l}}(\bx)$\footnote{{\blc As each AP serves all users, only one variable $\bx$ should be considered in the message passing process. Therefore, we have a reduced form as $\mu_{f_{l}\rightarrow \bx}(\bx) \propto \int f_{l}(\bx)\Pi \mu_{\bx \rightarrow f_{l}} \rmd\bx \propto   b_{f_{l}}(\bx)/\mu_{\bx \rightarrow f_l}(\bx)$.}}.} The message $\mu_{\bx \rightarrow f_{l}}(\bx)$ is initialized as
$\mathcal{N}_{\mathbb{C}}(\bx; \boldsymbol{\gamma}_{l}^{(0)}/\lambda_{l}^{(0)}, \lambda_{l}\bI)$ for each AP. Therefore, how to compute the belief $b_{f_{l}}(\bx)$ is of great significance and it is given by
\begin{equation}\label{eq_KL}
b_{f_l}(\bx)=\arg\min\limits_{b(\bx)\in\cF}D\left[ {{\mu_{\bx \to {f_l}}}\left( \bx \right){f_l}\left( \bx \right)}\|b(\bx) \right],
\end{equation}
The explicit expression for KL divergence is given in (\ref{KL_Definition}). Since obtaining a distribution to minimize the KL divergence is very difficult, we consider $b_{f_l}(\bx)$ is a Gaussian distribution with the same mean and covariance matrix with the distribution $\mu _{\bx \to {f_l}}\left( \bx \right){f_l}\left( \bx \right)$.  Therefore, we have
{\blc $b_{f_l}(\bx) = \mathcal{N}_{\mathbb{C}}(\bx; \boldsymbol{\mu}_{l}, \boldsymbol{\Sigma}_{l})$ with $\boldsymbol{\Sigma}_{l} = (\sigma^{-2} \bH^{H}_{l}\bH_{l}+\lambda_{l}\bI)^{-1} $and $\boldsymbol{\mu}_{l} = \boldsymbol{\Sigma}_{l} (\sigma^{-2}\bH_{l}\by_{l}+
\boldsymbol{\gamma}_{l})$,} which  yields (\ref{eqvarA}) and (\ref{eqmeanA}) in \textbf{Algorithm 1}. Then, we have
\begin{equation}\label{eqmessage}
  \mu_{f_l\rightarrow \bx}(\bx) = \frac{b_{f_l}(\bx)}{\mu_{\bx\to {f_l}}(\bx)} \cong \frac{\mathcal{N}_{\mathbb{C}}(\bx; \bx_{\rmA,l}^{\mathrm{post}}, v_{\rmA,l}^{ \mathrm{post}}\bI)}{\mathcal{N}_{\mathbb{C}}(\bx; \boldsymbol{\gamma}_{l}/\lambda_{l}, \lambda_{l}\bI)} \propto {\mathcal{N}_{\mathbb{C}}(\bx; \bx_{\rmA,l}^{\mathrm{ext}}, v_{\rmA,l}^{\mathrm{ext}}\bI)},
\end{equation}
{\blc where we approximate $b_{f_l}(\bx)$ by $\mathcal{N}_{\mathbb{C}}(\bx; \bx_{\rmA,l}^{\mathrm{post}}, v_{\rmA,l}^{ \mathrm{post}}\bI)$. The principles behind the approximation are \emph{moment-matching} and \emph{self-averaging} \cite{EP1,VAMP}. According to the Gaussian production lemma, we have (\ref{eqextvar}) and  (\ref{eqextmean}) to obtain
$v_{\rmA,l}^{\mathrm{ext}}$ and $\bx_{\rmA,l}^{\mathrm{ext}}$ in \textbf{Algorithm 1}, respectively.} Next, we consider the message $\mu_{\bx \rightarrow f}(\bx)$ to be the product of messages passing from each APs (factors)$\{f_1,f_{2},...,f_{L}\}$ to $\bx$ and we have
\begin{equation}\label{eqmessage2}
  \mu_{\bx \rightarrow f}(\bx) = \prod_{l=1}^{L}\mu_{\bx \to {f_l}}= \prod_{l=1}^{L}\mathcal{N}_{\mathbb{C}}(\bx; \bx_{\rmA,l}^{\mathrm{ext}},
  v_{\rmA,l}^{\mathrm{ext}}) = \mathcal{N}_{\mathbb{C}}(\bx; \bx_{\rmA}^{\mathrm{ext}}, v_{\rmA}^{\mathrm{ext}}\bI),
\end{equation}
where {\blc $v_{\rmA}^{\mathrm{ext}}$ and $\bx_{\rmA}^{\mathrm{ext}}$} are obtained from (\ref{eqvarmrc}) and (\ref{eqmeanmrc}). We then approximate the belief on $f$ as $b_{f}(\bx) \propto \mathcal{N}_{\mathbb{C}}(\bx; \bx_{\rmB}^{\mathrm{post}}, \mathrm{diag}(\bv_{\rmB}^{\mathrm{post}}))$. The mean $\bx_{\rmB}^{\mathrm{post}}$ and variance $\bv_{\rmB}^{\mathrm{post}}$ are obtained by \emph{moment-matching} and \emph{self-averaging}, and denoted by (\ref{eqmeanB}) and (\ref{eqvarB}), respectively. As the transmitted symbol is assumed to be drawn from $M$-QAM set, the explicit expressions for (\ref{eqmeanB}) and (\ref{eqvarB}) are given by (\ref{eqmean}) and (\ref{eqvar}). Similar to (\ref{eqmessage}), we set the message $\mu_{f \rightarrow \bx}(\bx)$ as

\begin{equation}\label{eqmessage3}
  \mu_{f \rightarrow \bx}(\bx) = \frac{b_{f}(\bx)}{\mu_{\bx \rightarrow f}(\bx)} = \frac{\mathcal{N}_{\mathbb{C}}(\bx; \bx_{\rmB}^{\mathrm{post}}, v_{\rmB}^{\mathrm{post}} \bI)}{\mathcal{N}_{\mathbb{C}}(\bx; \bx_{\rmA}^{\mathrm{ext}}, v_{\rmA}^{\mathrm{ext}}\bI)} \propto
   \mathcal{N}_{\mathbb{C}}(\bx,  \bx_{\rmB}^{\mathrm{ext}}, v_{\rmB}^{\mathrm{ext}}).
\end{equation}
Finally, according to the variable-to-factor messages, we compute the message $\mu_{\bx \rightarrow f_{l}}(\bx)$ for the next iteration which is given by
\begin{equation}\label{eqmessage4}
  \mu_{\bx \rightarrow f_{l}}(\bx) = \mu_{f \rightarrow \bx }(\bx) \prod_{f^{'}\neq f} \mu_{f^{'} \rightarrow \bx }(\bx)
  \propto \mathcal{N}_{\mathbb{C}}(\bx; \boldsymbol{\gamma}_{l}/\lambda_{l}, \lambda_{l}\bI),
\end{equation}
and the explicit expressions for $\boldsymbol{\gamma}_{l}$ and $\lambda_{l}$ are given by (\ref{eqvarest}) and (\ref{eqmmeanest}), respectively. We can obtain the \textbf{Algorithm 1} by repeating above  message passing procedure.

{\blc
\subsection{Derivation of MRC Combining} \label{derivation33}

For distributed EP-based MIMO detection, the challenge is how to combine the message from different APs. In our paper, we focus on linear combining of the local estimate $\mathbf{x}_{\mathrm{A},l}^{\mathrm{ext}}$ and  $v_{\mathrm{A},l}^{\mathrm{ext}}$ from APs as follows,
\begin{equation}\label{eqmeanmrcA1}
    \mathbf{x}_{\mathrm{A}}^{\mathrm{ext}} = \sum_{l=1}^{L}a_{l}\mathbf{x}_{\mathrm{A},l}^{\mathrm{ext}}, \quad l = 1,\ldots,L .
\end{equation}

As the linear model ($\ref{eqyd}$)   is decoupled into the equivalent AWGN model, we focus on the scalar form of the ($\ref{eqmeanmrcA1}$) with
\begin{equation}\label{eqmeanmrc_scalar}
    x_{k,\mathrm{A}}^{\mathrm{ext}} = \sum_{l=1}^{L}a_{l,k}x_{\mathrm{A},l,k}^{\mathrm{ext}}, \quad k = 1,\ldots,K,
\end{equation}
where $a_{l,k}$ is the local estimate for the $k$-user from the $l$-th AP, and it depends on the variance $v_{\mathrm{A},l}^{\mathrm{ext}}$. As in ($\ref{eqAWGNl}$) and ($\ref{eqAWGN}$)., we write the estimate $x_{k,\mathrm{A}}^{\mathrm{ext}} = x_{k}+ n_{l,k}^{eq}$, where $n_{l,k}^{eq}$ represents equivalent noise with known variance  $v_{\mathrm{A},l}^{\mathrm{ext}}$. For the $k$-th user, the optimal linear combining is $x_{k,\mathrm{A}}^{\mathrm{ext}} = \sum_{l=1}^{L}a_{l,k}x_{\mathrm{A},l,k}^{\mathrm{ext}}$. Thus, we have $x_{k,\mathrm{A}}^{\mathrm{ext}} = \sum_{l=1}^{L}a_{l,k}x_{k}+\sum_{l=1}^{L}a_{l,k}n_{\mathrm{A},l,k}^{\mathrm{ext}}$. Because $x_{k,\mathrm{A}}^{\mathrm{ext}}$ should be an unbiased estimate of $x_{k}$, we have $\sum_{l=1}^{L}a_{l,k}=1$. After combining, the equivalent SNR is given by
\begin{equation}\label{SNR_Combine}
    \mathrm{SNR} = \frac{|x_{k}|^{2}}{|\sum_{l=1}^{L}a_{l,k}n_{\mathrm{A},l,k}^{\mathrm{ext}}|^{2}} = \frac{E_s}{\sum_{l=1}^{L}a_{l,k}^{2}v_{\mathrm{A},l,k}^{\mathrm{ext}}}.
\end{equation}
As a result, the combing problem becomes an optimization problem
\begin{align}\label{eqopt}
  \max  & \quad \frac{E_s}{\sum_{l=1}^{L}a_{l,k}^{2}v_{\mathrm{A},l,k}^{\mathrm{ext}}} \nonumber \\
  \mathrm{s.t.} & \quad \sum_{l=1}^{L}a_{l,k}=1.
\end{align}
By using the method of Lagrange multipliers, it is easy to obtain
\begin{equation}\label{eqmeanmrc_scalar}
    a_{k,l} = \frac{1}{v_{\mathrm{A},l,k}}\bigg(\sum_{l=1}^{L} \frac{1}{ v_{\mathrm{A},l,k}^{\mathrm{ext}}}\bigg)^{-1}, l = 1,\ldots,L.
\end{equation}
Thus, we obtain the MRC combining in ($\ref{eqvarmrc}$) and ($\ref{eqmeanmrc}$).
}

\section{Derivation of ICD} \label{derivation2}
In this appendix, we present the derivation of the equivalent channel estimation error in the data detection stage, as well as
the derivation of data detection error in channel estimation stage.
\subsection{Derivation of Equivalent Channel Estimation Error} \label{derivation21}
{\blc The equivalent noise $\hat{\bn}^{\rmd}[n]$ in the signal detector }
is assumed to be Gaussian distributed with zero mean and 
covariance matrix $\hat{\bV}_{\mathrm{est}}[n]$, which can be obtained by considering the statistical properties of the channel estimation error. For the convenience of simplicity, we omit the time index $n$. The detailed expression for $\hat{\mathbf{V}}_{\mathrm{est}}$ is then given by
\begin{equation}\label{covvd}
  \hat{\mathbf{V}}_{\mathrm{est}} = \mathrm{diag} \left(\sum_{j=1}^{K}\sigma_{\Delta h_{1,j}}^{2}+\sigma^{2},\ldots,\sum_{j=1}^{K}\sigma_{\Delta h_{N,j}}^{2}+\sigma^{2} \right),
\end{equation}
where $\sigma_{\Delta h_{i,j}}^{2}$ is the variance of $(i,j)$-th element for  channel estimation error matrix $\Delta\bH$ obtained from $\bR_{\Delta{\bh_l^\rmp}}$.
\subsection{Derivation of Data Detection Error} \label{derivation22}
Owing to the data feedback from the signal detector to the channel estimator, the LMMSE channel estimation in  the data-aided stage is given by
\begin{equation}\label{LMMSE2}
 \hat{\bar{\bh}}_{l} = \bR_{\bar{\bh}_{l} \bar{\bh}_{l}}\bA^{H}(\bA \bR_{\bar{\bh}_{l} \bar{\bh}_{l}} \bA^{H}+\bR_{\bn \bn})^{-1}\by,
\end{equation}
where  $\bA = \bX^{T} \otimes \mathbf{I}_{N} \in \mathbb{C}^{\tau_{c} N \times K N}$,  $\by=\mathrm{vec}(\mathbf{Y}) \in \mathbb{C}^{\tau_c N \times 1}$,  $ \bn= \mathrm{vec}(\mathbf{N}) \in \mathbb{C}^{\tau_{c}N \times 1} $, and $ \tau_c = \tau_p + \tau_d$. The covariance matrix
$ \bR_{\Delta\bH_{l}}$  of the  channel estimation error vector $\Delta{\bh_l}=\hat{\bar{\bh}}_l-\bh$ can be computed as
\begin{equation}\label{LMMSE_error2}
 \bR_{\Delta \bh_l} = \bR_{\bar{\bh}_{l} \bar{\bh}_{l}}-\bR_{\bar{\bh}_{l} \bar{\bh}_{l}}\bA^{H}(\bA \bR_{\bar{\bh}_{l} \bar{\bh}_{l}} \bA^{H}+\bR_{\bn \bn})^{-1}\bA\bR_{\bar{\bh}_{l} \bar{\bh}_{l}}.
\end{equation}
The covariance matrix $\bR_{\bn \bn}$ contains the equivalent noise power of the actual pilot $\mathbf{X}^{\rmp}$ and additional pilot  $ \hat{\mathbf{X}}^{\rmd}$. The explicit expression for $\bR_{\bn \bn}$ is given by
\begin{align}\label{finalSVD}
\bR_{\bn \bn} = \left[\begin{array}{c c c}
 \sigma^{2}\bI_{\tau_p N}  \\
   & \hat{\mathbf{V}}_{\mathrm{det}}
\end{array}\right],
\end{align}
where
\begin{align}\label{finalSVD}
 \hat{\mathbf{V}}_{\mathrm{det}} = \left[\begin{array}{c c c}
 \hat{\mathbf{V}}^{\rmd}[1]  \\
 &\ddots   \\
 &&\hat{\mathbf{V}}^{\rmd}[{\tau_d}]
\end{array}\right]
\end{align}
and
\begin{equation}\label{eqvp}
  \hat{\mathbf{V}}^{\rmd}[n] = \left(\sum_{j=1}^{K_{l}}\sigma_{e_{j,n}}^{2}+\sigma^{2}\right)\mathbf{I}_{K}.
\end{equation}
Denote $\bz_{n}=\mathbf{H}\bee_{n}$, $z_{i,n} = \sum_{j=1}^{K}h_{i,j}e_{j,n}$,  where $\bee_{n}$ is the $n$-th column of the signal detection error matrix $\bE^{\rmd}$ and $\sigma_{e_{j,n}}^{2}$ is the variance for $j$-th element of $\bee_{n}$. The signal detection error matrix $\bE^{\rmd}$ can be obtained from the posterior variance $\bv_{\rmA,l}^{\mathrm{post}}$ in the module A.

\end{document}